\documentclass[pdflatex,sn-mathphys-num]{sn-jnl}

\usepackage{graphicx}%
\usepackage{multirow}%
\usepackage{amsmath,amssymb,amsfonts}%
\usepackage{amsthm}%
\usepackage{mathrsfs}%
\usepackage[title]{appendix}%
\usepackage{textcomp}%
\usepackage{manyfoot}%
\usepackage{booktabs}%
\usepackage{algorithm}%
\usepackage{algorithmicx}%
\usepackage{algpseudocode}%
\usepackage{listings}%

\usepackage{todonotes}

\usepackage[T1]{fontenc}
\usepackage{inconsolata}

\theoremstyle{thmstyleone}%
\theoremstyle{thmstyletwo}%

\theoremstyle{thmstylethree}%

\usepackage{enumitem}
\usepackage{subcaption}
\usepackage[utf8]{inputenc}

\DeclareFixedFont{\ttb}{T1}{phv}{bx}{n}{9} 
\DeclareFixedFont{\ttm}{T1}{phv}{m}{n}{9}  

\usepackage{color}
\definecolor{deepblue}{rgb}{0,0,0.5}
\definecolor{deepred}{rgb}{0.6,0,0}
\definecolor{deepgreen}{rgb}{0,0.5,0}

\definecolor{mbdblue}{rgb}{0,0,0.5}
\definecolor{mbdyellow}{rgb}{0.6,0,0}
\definecolor{mbdpink}{rgb}{0.94509804, 0.61960784, 0.69411765}
\definecolor{mbdpinkdark}{HTML}{E65173}

\definecolor{deepgreen}{rgb}{0,0.5,0}

\definecolor{mbddark}{rgb}{0.55294118, 0.7254902 , 0.89803922}
\definecolor{rwthblue}{rgb}{0.0 , 0.32941176, 0.62352941}

\newcommand{\udarcy}{\mathbf{u}}
\newcommand{\pdarcy}{p}

\definecolor{halfgray}{gray}{0.55}
\definecolor{ipython_frame}{RGB}{207, 207, 207}
\definecolor{ipython_bg}{RGB}{247, 247, 247}
\definecolor{ipython_red}{RGB}{186, 33, 33}
\definecolor{ipython_green}{RGB}{0, 128, 0}
\definecolor{ipython_cyan}{RGB}{64, 128, 128}
\definecolor{ipython_purple}{RGB}{170, 34, 255}

\def\code#1{\texttt{#1}}
\usepackage{listings}

\newcommand\pythonstyle{\lstset{
language=Python,
basicstyle=\ttm,
morekeywords={access,and,break,class,continue,def,del,elif,else,except,exec,finally,for,from,global,if,import,in,is,lambda,not,or,pass,print,raise,return,try,while, as},%
morekeywords=[2]{abs,all,any,basestring,bin,bool,bytearray,callable,chr,classmethod,cmp,compile,complex,delattr,dict,dir,divmod,enumerate,eval,execfile,file,filter,float,format,frozenset,getattr,globals,hasattr,hash,help,hex,id,input,int,isinstance,issubclass,iter,len,list,locals,long,map,max,memoryview,min,next,object,oct,open,ord,pow,property,range,raw_input,reduce,reload,repr,reversed,round,set,setattr,slice,sorted,staticmethod,str,sum,super,tuple,type,unichr,unicode,vars,xrange,zip,apply,buffer,coerce,intern},%
sensitive=true,%
morecomment=[l]\#,%
morestring=[b]',%
morestring=[b]",%
morestring=[s]{'''}{'''},
morestring=[s]{"""}{"""},
morestring=[s]{r'}{'},
morestring=[s]{r"}{"},%
morestring=[s]{r'''}{'''},%
morestring=[s]{r"""}{"""},%
morestring=[s]{u'}{'},
morestring=[s]{u"}{"},%
morestring=[s]{u'''}{'''},%
morestring=[s]{u"""}{"""},%
literate=
*{+}{{{\color{ipython_purple}+}}}1
{-}{{{\color{ipython_purple}-}}}1
{*}{{{\color{ipython_purple}$^\ast$}}}1
{/}{{{\color{ipython_purple}/}}}1
{^}{{{\color{ipython_purple}\^{}}}}1
{?}{{{\color{ipython_purple}?}}}1
{!}{{{\color{ipython_purple}!}}}1
{\%}{{{\color{ipython_purple}\%}}}1
{<}{{{\color{ipython_purple}<}}}1
{>}{{{\color{ipython_purple}>}}}1
{|}{{{\color{ipython_purple}|}}}1
{\&}{{{\color{ipython_purple}\&}}}1
{~}{{{\color{ipython_purple}~}}}1
{==}{{{\color{ipython_purple}==}}}2
{<=}{{{\color{ipython_purple}<=}}}2
{>=}{{{\color{ipython_purple}>=}}}2
{+=}{{{+=}}}2
{-=}{{{-=}}}2
{*=}{{{$^\ast$=}}}2
{/=}{{{/=}}}2,
literate=
{á}{{\'a}}1 {é}{{\'e}}1 {í}{{\'i}}1 {ó}{{\'o}}1 {ú}{{\'u}}1
{Á}{{\'A}}1 {É}{{\'E}}1 {Í}{{\'I}}1 {Ó}{{\'O}}1 {Ú}{{\'U}}1
{à}{{\`a}}1 {è}{{\`e}}1 {ì}{{\`i}}1 {ò}{{\`o}}1 {ù}{{\`u}}1
{À}{{\`A}}1 {È}{{\'E}}1 {Ì}{{\`I}}1 {Ò}{{\`O}}1 {Ù}{{\`U}}1
{ä}{{\"a}}1 {ë}{{\"e}}1 {ï}{{\"i}}1 {ö}{{\"o}}1 {ü}{{\"u}}1
{Ä}{{\"A}}1 {Ë}{{\"E}}1 {Ï}{{\"I}}1 {Ö}{{\"O}}1 {Ü}{{\"U}}1
{â}{{\^a}}1 {ê}{{\^e}}1 {î}{{\^i}}1 {ô}{{\^o}}1 {û}{{\^u}}1
{Â}{{\^A}}1 {Ê}{{\^E}}1 {Î}{{\^I}}1 {Ô}{{\^O}}1 {Û}{{\^U}}1
{œ}{{\oe}}1 {Œ}{{\OE}}1 {æ}{{\ae}}1 {Æ}{{\AE}}1 {ß}{{\ss}}1
{ç}{{\c c}}1 {Ç}{{\c C}}1 {ø}{{\o}}1 {å}{{\r a}}1 {Å}{{\r A}}1
{€}{{\EUR}}1 {£}{{\pounds}}1,
keywordstyle=\ttb\color{black},
emphstyle=\ttb\color{rwthblue},    
stringstyle=\color{rwthblue},
showstringspaces=false,
numbers=left,
numberstyle=\tiny\color{halfgray},
}}

\lstnewenvironment{python}[1][]
{
\pythonstyle
\lstset{#1}
}
{}

\newcommand\pythoninline[1]{{\pythonstyle\lstinline!#1!}}

\newcommand\YAMLcolonstyle{\color{mbdpinkdark}\ttm}
\newcommand\YAMLkeystyle{\color{black}\ttm}
\newcommand\YAMLvaluestyle{\color{rwthblue}\ttm}

\makeatletter

\newcommand\language@yaml{yaml}

\expandafter\expandafter\expandafter\lstdefinelanguage
\expandafter{\language@yaml}
{
  keywords={true,false,null,y,n},
  keywordstyle=\color{darkgray}\bfseries,
  basicstyle=\YAMLkeystyle,                                 
  sensitive=false,
  comment=[l]{\#},
  morecomment=[s]{/*}{*/},
  commentstyle=\color{purple}\ttm,
  stringstyle=\YAMLvaluestyle\ttm,
  moredelim=[l][\color{orange}]{\&},
  moredelim=[l][\color{magenta}]{*},
  moredelim=**[il][\YAMLcolonstyle{:}\YAMLvaluestyle]{:},   
  morestring=[b]',
  morestring=[b]",
  literate =    {---}{{\ProcessThreeDashes}}3
                {>}{{\textcolor{red}\textgreater}}1     
                {|}{{\textcolor{red}\textbar}}1 
                {\ -\ }{{\mdseries\ -\ }}3,
}

\lst@AddToHook{EveryLine}{\ifx\lst@language\language@yaml\YAMLkeystyle\fi}
\makeatother

\newcommand\ProcessThreeDashes{\llap{\color{rwthblue}\mdseries-{-}-}}

\begin{document}

\title[Bryne: sustainable prototyping of finite element models]{Bryne: sustainable prototyping of finite element models}


\author*[1]{\fnm{Benjamin} \sur{Terschanski}}\email{terschanski@mbd.rwth-aachen.de}
\author[2]{\fnm{Robert} \sur{Kl\"ofkorn}}\email{robertk@math.lu.se}
\author[3]{\fnm{Andreas} \sur{Dedner}}\email{a.s.dedner@warwick.ac.uk}
\author*[1]{\fnm{Julia} \sur{Kowalski}}\email{kowalski@mbd.rwth-aachen.de}


\affil[1]{\orgdiv{Chair of Methods for Model-based Development in Computational Engineering}, \orgname{Faculty of Mechanical Engineering, RWTH Aachen}, \orgaddress{\street{Eilfschornsteinstraße 18}, \city{Aachen}, \postcode{52062}, \country{Germany}}}

\affil[2]{\orgname{Center for Mathematical Sciences, Lund University}, \orgaddress{\street{Box 117}, \city{Lund}, \postcode{221 00}, \country{Sweden}}}

\affil[3]{\orgname{Mathematics Institute, University of Warwick}, \orgaddress{\street{Coventry}, \city{Warwick}, \postcode{CV4 7AL}, \country{United Kingdom}}}


\abstract{
  Open-source simulation frameworks are evolving rapidly to provide accessible tools for the numerical solution of partial differential equations.
  Modern finite element (FEM) software such as FEniCS, Firedrake, or dune-fem alleviates the need for modelers to recode the discretization and linear solver backend for each application and enables rapid prototyping of solvers.
  However, while it has become easier to build prototype FEM models, creating a solver reusable beyond its specific initial simulation setup remains difficult. 
  Moreover, simulation setups typically cover an ample input parameter space, and tracking complex metadata on research project time scales has become a challenge.
  This implies the need to supplement model development with a coding-intensive complementary workstream, seldom developed for sustainable reuse.  
  To address these issues, we introduce our open-source Python package Bryne.

  Bryne is an object-oriented framework for FEM solvers built with the dune-fem Python API.
  In this article, we describe how it helps to evolve rapid-prototyping solver development into sustainable simulation building.
  First, we show how to translate a minimal dune-fem solver into a Bryne FEM model to build human-readable, metadata-enriched simulations. 
  Bryne then offers a simulation driver and model coupling interfaces to combine implemented solvers in operator-split multiphysics simulations.
  The resulting reproducibility-enabled infrastructure allows users to tackle complex simulation setups without sacrificing backend flexibility.
  We demonstrate the workflow on a convection-coupled phase-change simulation, where a discontinuous Galerkin flow solver is coupled with a solver for solidification phase change.
}


\keywords{research software engineering, multiphysics, reproducible science, finite element method, solidification phase change}



\maketitle
\section{Introduction}\label{sec1}
Modern open-source numerics frameworks provide modelers with powerful tools to build complex simulations.
For simulations based on finite element (FEM) discretization a variety of open-source codes such as dealII \cite{dealii2019design}, elmer \cite{kondovMultiscaleModellingMethods2013}, MFEM \cite{kondovMultiscaleModellingMethods2013}, FEniCS \cite{barattaDOLFINxNextGeneration}, Firedrake \cite{rathgeberFiredrakeAutomatingFinite2017}, dune-fem \cite{bastianDuneFrameworkBasic2021}, \cite{dednerPythonBindingsDUNEFEM2020a} are under active development, many of which provide a Python API.

Notably, Fenics, Firedrake, and dune-fem allow users to input the weak form of the partial differential equation (PDE), the Unified Form Language (UFL) \cite{alnaesUnifiedFormLanguage2014a}.
Alleviating the need to re-implement discretization and linear solvers, they allow for rapid prototyping of models and numerical methods and shift the focus from low-level debugging to numerical modeling.
The respective backend then handles the assembly of the linear system and the operator Jacobian via automatic code generation. 
For performance reasons, the linear solves are typically performed in a lower-level language such as C++.
This approach allows for easy experimentation with fundamental simulation components, such as different approximation spaces, solver setups, preconditioners, or even completely new PDE terms.
Subsequently, Python-API-driven open-source codes had a significant impact on the computational science community. \\

While the advent of Python API-based FEM solvers has facilitated initial prototype creation, it inadvertently created a new reproducibility challenge. 
Here, we introduce the term "sustainable prototyping" to describe the process of moving from a draft solver to a piece of numerical software with two re-user perspectives in mind: 
\begin{itemize}
  \item For the simulation model developer, the code that implements solver for an FEM model has to be maintainable and extensible.
  \item For users and other researchers, simulation setups should be fully transparent in a way that does not require deep knowledge of the solver backend.
\end{itemize}

\begin{figure}
  \centering
  \includegraphics[width=\textwidth]{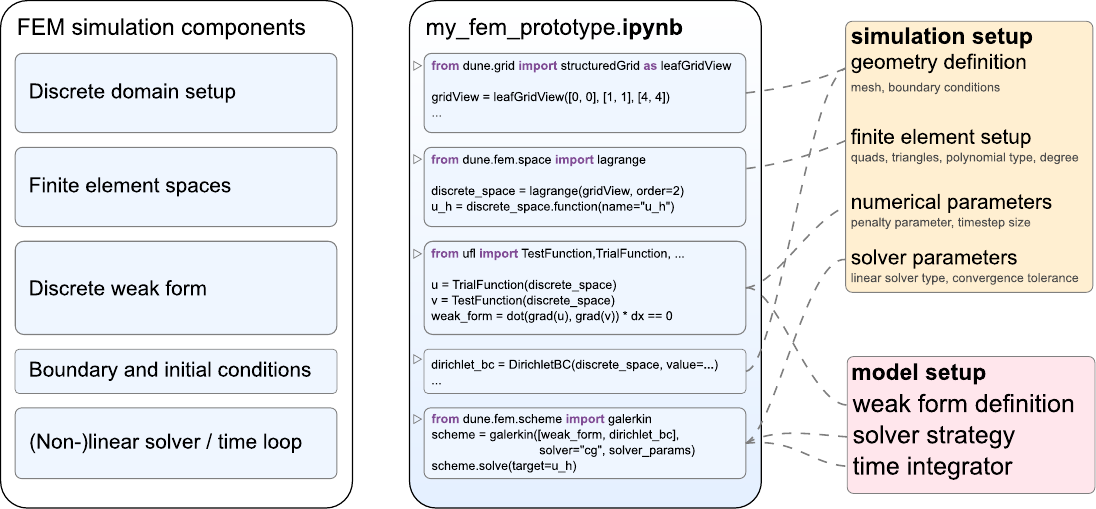}
  \caption[Sketch of a typical Python FEM simulation notebook]{Sketch of a typical Python FEM simulation notebook (\textit{center}). In a prototyping workflow, users implement the core components of a simulation (\textit{left}) in a single Python module or notebook (\textit{center}). On the right, we distinguish between the \textit{simulation setup}, containing the input parameters or "moving parts" of the simulation, and the \textit{model setup}, which is invariant for a given model.}
  \label{fig:intro_notebooks}
\end{figure}

In practice, there is often a trade-off between the flexibility in rapid-prototyping of new PDE solvers and the degree of sophistication of the simulation setup infrastructure. 
Figure \ref{fig:intro_notebooks} illustrates a typical monolithic prototype solver. 
Here, all major simulation components are implemented in a single notebook or Python module.
This includes the static model definition and solver logic (\textit{model setup}), as well as the inputs specific to a particular simulation scenario (\textit{simulation setup}). 
The specific simulation setup includes the mesh type, finite element basis and degree, physical coefficients, material parameters, as well as numerical and solver parameters.
The input parameter space can thus become quite large even for a minimal, single-purpose PDE solver.   

Prototype models and showcase solvers like the one in Fig. \ref{fig:intro_notebooks} are reproducible as functional early samples to prove the feasibility of a simulation method.  
Reusability is, however, limited in several ways:
\begin{enumerate}
  \item A tight coupling of solver logic and simulation setup sacrifices flexibility in changing the application scenario.
  \item Models can not easily be generalized or combined in multi-physics simulations. 
  \item Non-standard interfaces and input handling make it hard to compare and archive models and simulation results.
\end{enumerate}
It hence remains challenging to develop simulation codes on top of popular open-source numerical backends that are usable past an initial proof of concept. \\

Some efforts have been made to address these issues. 
Probably the most sophisticated project is the commercial software FEATool Multiphysics \cite{FEATool2025}, which implements a simulation builder using the FEniCS backend, including a GUI.
The design of FEATool integrates some of the concepts discussed in this paper, but it is not free software and the base paid version does not include the capabilities for user-defined models.
Community efforts such as the FEniCS component multiphenicsx \cite{multiphenicsx} focus on the implementation of solver logic for coupled multiphysics simulations, but do not provide a higher-level framework for building reusable FEM models. To the best of our knowledge, no such efforts exist for Firedrake and dune-fem at the time of writing.
Subsequently, application developers of open-source research solvers built on FEM backends are forced to re-implement their own simulation infrastructure, hindering the establishment of community best practices for the reproducibility of complex simulation setups. Recent examples can be found using Fenics \cite{bergersenTurtleFSIRobustMonolithic2020}, \cite{ogucFeVAcSPackageVisualizing2025}, \cite{quintinoOpytimalPythonFEniCS2025}, \cite{wintzerMultiphysicsSimulationCrystal2025}, Firedrake \cite{kjeldsbergVerifiedValidatedMoving2023}, \cite{kyasAcceleratedReactiveTransport2022} and dune-fem \cite{dednerPythonFrameworkHpadaptive2019a}, \cite{terschanskiStableRegimesMixed2025}. \\

We therefore identify an infrastructural gap between the Python-API-based discretization tools that enable flexible solver building and the requirements for the sustainable development of complex simulations.
Bridging this gap is a core motivation of this work. 
In this paper, we present Bryne, a Python package built on the dune-fem Python API and designed to
\begin{itemize}
  \item semantically structure and systematically compose prototyped models in a modular way, and to
  \item improve metadata traceability and hence the reproducibility of complex simulation setups.
\end{itemize}
Bryne wraps the flexibility of UFL and dune-fem in a solver-building architecture that allows users to move from first prototypes to reusable models. \\

This is achieved by a modular design, separating the simulation setup from the model definition and solver logic (\textit{see right column of Fig. \ref{fig:intro_notebooks}}). 
In Bryne, the definition of a finite element model comprises its weak form, a definition of model-specific parameters, and the solver logic to compute a numerical solution to the model governing PDE.
Optionally, the model can also provide a coupling interface to other models, easing data exchange between solvers on the same discrete domain. 
Bryne has an object-oriented structure that allows for extension of existing models through inheritance.
The package handles simulation inputs through YAML files and user-modifiable Python scripts imported from a directory at runtime.
While containing the complete simulation setup, this input directory aims to be easy to navigate and doubles as a human-readable documentation of the simulation setup. To this end, the simulation results are always automatically enriched with the setup metadata, which enables easy parameter traceability and reproduction. To our knowledge, this is the first such open-source package for building reusable FEM models with dune-fem. \\

It should be noted that Bryne does not implement any low-level finite element assembly or linear solver and is, therefore, not a standalone solver library. 
In the present version of Bryne, we rely on the dune-fem package for the finite element discretization and linear solvers. 
Dune-fem in turn supports PETSc as a linear solver backend and has an interface to petsc4py \cite{petsc4py:11}, \cite{ petsc3_22}.
Weak form definitions in dune-fem are based on the same UFL syntax as other open-source libraries, and many of the concepts presented in the following will translate to other FEM backends with Python bindings.
A minimal FEniCS or Firedrake solver will look very similar to Fig. \ref{fig:intro_notebooks}, thus it becomes relatively straightforward to extend the Bryne architecture to include models written in these libraries.
Hence, this work's contribution also lies in the proposition of general simulator components to elevate the general usability of Python API discretization backends. \\

All major software components are described in Section \ref{sec:bryne_software_design}, starting with an example application scenario to motivate our design requirements. 
We spend some time explaining the building blocks of Bryne FEM models. 
A common interface enables model extension through inheritance and data exchange of models on the same grid; this value-add is demonstrated using an example model hierarchy.

In Section \ref{sec:application_examples}, we demonstrate Bryne usage in a research context.
We combine a discontinuous Galerkin flow solver with a thermal phase-change solver to build a convection-coupled phase change simulation. 
Verification results and the solver codes are first published in this article.

\section{Bryne software design} \label{sec:bryne_software_design}

In this section, we introduce major components of the Bryne software package and how they interact to build a simulation framework on top of the dune-fem Python API. 

\subsection{Motivational example}

To understand the scope of the present work, we start with a simulation example inspired by our research. It serves as a "modeler's perspective" motivation for the development of Bryne and doubles as an introduction to the software design. \\

\begin{figure}
  \centering
  \includegraphics[width=\textwidth]{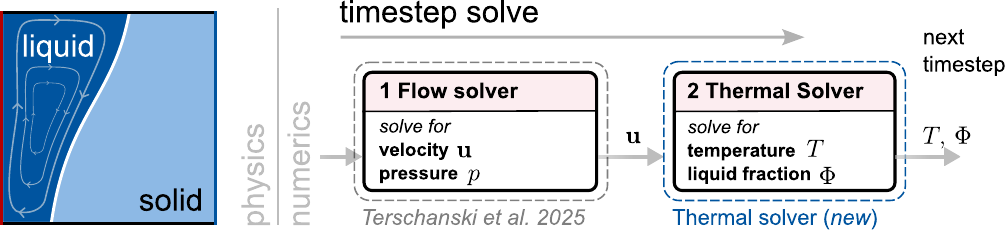}
  \caption[Intro example: convection-coupled flow]{
    Phase change problem of a solid melting into a fluid moving due to natural convection (\textit{left}). 
    The sketch on the right shows an example numerical model sequence of a flow solver (\textit{1}) and a two-phase thermal solver (\textit{2}) executed in sequentially in an operator-split fashion.
  }
  \label{fig:intro_example}
\end{figure}

Imagine we would like to simulate a convection-coupled phase change process, where a solid exposed to a heat source is progressively melting, while the melt is moving due to natural convection. 
The setup is sketched in Fig \ref{fig:intro_example}, together with a common numerical approach to solve the multiphysics problem in an operator-split fashion with flow and temperature fields solved sequentially \cite{parkinsonModellingBinaryAlloy2020}, \cite{kaaksEnergyconservativeDGFEMApproach2023}. 
To build the final complex process model, we would have to   
\begin{enumerate}
  \item develop the individual model components, e.g. the flow solver (\textit{(1) in Fig. \ref{fig:intro_example}}) to compute velocity and pressure, and the two-phase thermal solver \textit{(2)} to compute the temperature and liquid volume fraction. 
  \item perform incremental quality assurance for both models, such as verification against analytical and manufactured solutions as well as empirical convergence studies.
  \item develop a multi-physics coupling strategy that combines individual model components into a multi-physics solver.
  \item embed the resulting multi-physics model into a simulation driver that will allow us to either study the physical process or to validate against data to test admissibility of the model's assumptions.
\end{enumerate}
At all points, we want to be able to experiment with different finite element methods, linear solvers, and mesh types. 
To perform validation and to study real physical processes, we need to keep track of increasingly complex simulation setups, including material parameters for both phases and numerical parameters for two individual non-trivial finite element solvers.
Finally, we would like to publish our results. 
To build trust in our method and enable independent reproduction, we must share simulation setup metadata with reviewers and readers. \\

Building upon the flexibility of the dune-fem Python API, Bryne provides a single framework to achieve these tasks.

\subsection{Overview of software components}

Bryne simulations rest on four pillars, illustrated in Figure \ref{fig:bryne_overview}:  

\begin{figure}
  \centering
  \includegraphics[width=\textwidth]{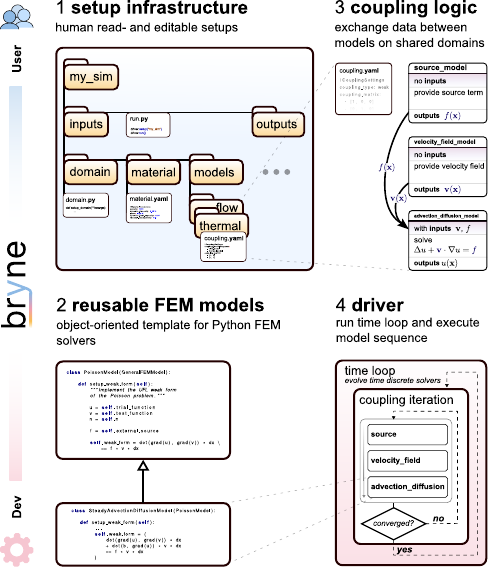}
  \caption[Core components of the Bryne software package]{An overview of core components of the Bryne software package. Our package provides four main contributions on top of the dune-fem Python API.
  A logical split between the core implementation and simulation allows for simulation setups in a semantically structured directory, which can be stored with results to ensure metadata traceability (\textit{1}).
  Simulations are built from model solvers, which are implemented as classes with a common interface (\textit{2}). 
  Subsequently, developers can declare model inputs and outputs, which allows models on the same grid to exchange data (\textit{3}).
  The Bryne main driver provides a common timestepping routine and executes the model solver sequence in a simulation loop (\textit{4}).
  }
  \label{fig:bryne_overview}
\end{figure}

\begin{enumerate}
  \item The Bryne \textbf{setup infrastructure}, input and output management introduces a semantically structured input directory. 
  Simulation setups are stored with the results, allowing for re-execution of the exact setup with Bryne and documentation of metadata for external reproduction. 
  \item Simulations are composed of \textbf{FEM models} implemented as Python classes sharing a common interface. 
  Each model implements the solver to a particular PDE problem.
  The object-oriented approach allows extending models through inheritance.
  \item The Bryne \textbf{driver} runs the main simulation time loop. 
  At each time step, it executes a sequence of models in a user-defined coupling sequence. 
  \item The \textbf{coupling logic} allows models defined on the same discrete grid to exchange data. 
  This can be used to combine models in an operator-split fashion.
\end{enumerate}

At the heart of Bryne lies the implementation of FEM models as Python classes, and we dedicate some space to convey the building blocks of a Bryne model. 

\subsection{Building models with Bryne}

\textit{All following minimal working examples are shipped with the Bryne repository \cite{bryne_software_repo}. Executable simulation setups can be found in the \code{examples} directory. 
Code snippets are excerpts of the respective example files; some are shortened for conciseness.} \\ 

Bryne simulations are built from reusable components, implementing mathematical model solvers as Python classes.
In the current version of Bryne, we focus on numerical PDE solvers based on finite element discretizations. 
In this section, we focus on two aspects of reusability, 
\begin{enumerate}
  \item \textbf{inheritance}, which allows for efficient extension of existing models and 
  \item  model \textbf{coupling} to build complex simulations from individually developed and tested solvers.
\end{enumerate}

\begin{figure}
  \centering
  \includegraphics[width=\textwidth]{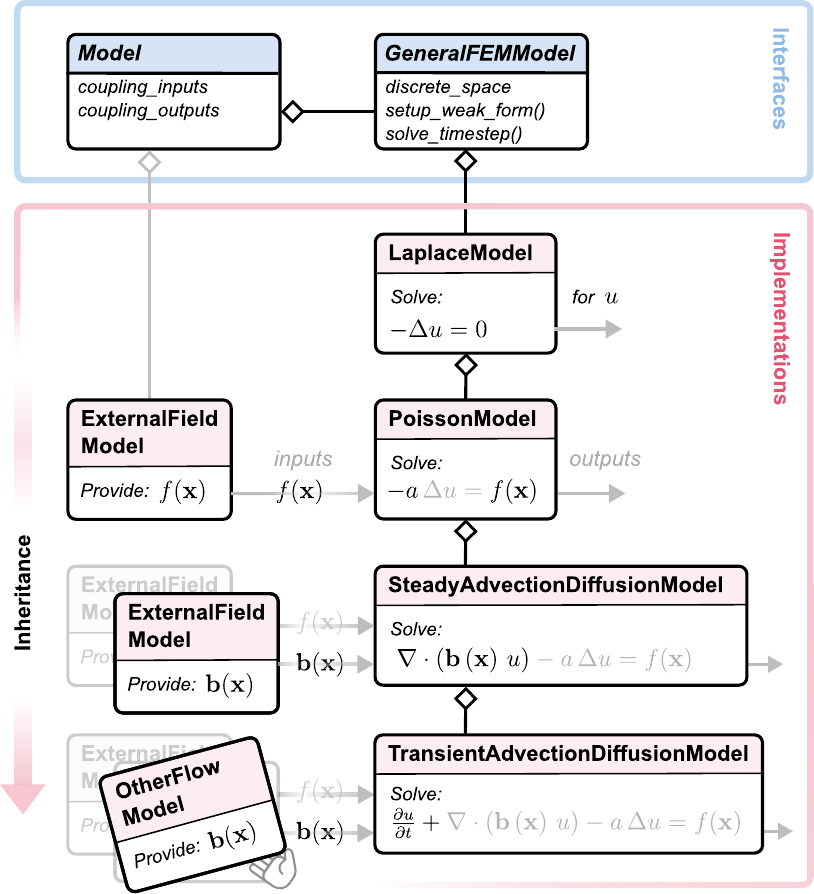}
  \caption[Solver cascade through inheritance]{
  Sketch of a model cascade through inheritance. Bryne FEM solvers are implemented as classes with a shared interface. 
  Inheritance allows building upon existing models by extending the weak form implementation, coupling interface, and solver logic. 
  The figure sketches the structure of model building examples presented in this article from top to bottom.
  Starting with the \code{LaplaceModel}, each row shows one example, where the flow of data through the Bryne coupling interface is indicated from left to right. 
  }
  
  \label{fig:coupling_inheritance}
\end{figure}

Figure \ref{fig:coupling_inheritance} illustrates the model hierarchy discussed in this section. Starting from the abstract \code{Model} and \code{GeneralFEMModel} classes, a sequence of increasingly complex models (\textit{top to bottom}) can be built by deriving from existing models. 
Each FEM model implements its own weak form definition but common terms, parameters and solver logic can all be inherited. 
Coupling between models on the same spatial grid can be realized by declaring solver inputs and outputs through a common coupling interface. 
This way, data can be exchanged in a coupling sequence (\textit{left to right}).  

\subsubsection{Basic concepts}


We start with a simple example to illustrate the idea.
Let's say we would like to solve the Laplace equation for scalar $u$, 
\begin{subequations}
\begin{align}\label{eq:laplace_equation}
  -\Delta u = 0 &  \quad \mathbf{x} \in \Omega & \quad ,\\
  u = g & \quad \mathbf{x} \in \partial\Omega & \quad ,
\end{align}
\end{subequations} 
where $\Omega \in \mathbb{R}^d$ is the $d-$dimensional domain and $g$ is a constant Dirichlet boundary condition on the entire boundary $\partial\Omega$. In Bryne, a finite element model for this PDE seeks to find the numerical approximation $u_h$ of the solution $u$ in a finite element space $V_h$. A core modeling step in building a Bryne FEM model, therefore, is the definition of a particular weak form of the PDE, which is then used to build and solve the linear system using the dune-fem backend \cite{dednerPythonBindingsDUNEFEM2020a}. 
In our first example, the classical variational problem reads
\begin{subequations}
\begin{align}
& \textit{continuous: } & & \text{Find }u\in V\text{ s.t. } & \int_{\Omega} \boldsymbol{\nabla} u \cdot \boldsymbol{\nabla} v \, d\Omega = 0\,,& \quad \forall v \in V_{0} & \quad , \\
& \textit{discrete: }& & \text{Find }u_h\in V_h\text{ s.t. } & \int_{\Omega_h} \boldsymbol{\nabla} u_h \cdot \boldsymbol{\nabla} v_h \, d\Omega_h = 0 \,, & \quad \forall v_h \in V_{h,0} & \quad , \label{eq:weak_form_laplace_discrete}
\end{align}
\end{subequations}
where $u$ and $v$ are continuous trial and test functions living in the spaces $V$ and $V_0$, respectively, where subscript $\bullet_0$ denotes that test functions vanish on the boundary $\partial\Omega$. 
We use subscript $\bullet_h$ to denote the discrete approximation to the continuous counterparts and hence $u_h$, $v_h$ are discrete trial and test functions.
Since we want to enable Bryne users to change the finite element space as part of the simulation setup, the discrete approximation space $V_h$ is deliberately left unspecified. 
To build our Laplace model solver in Bryne, we can now write a Python class \code{LaplaceModel} that defines 
\begin{enumerate}
  \item the discrete weak form \eqref{eq:weak_form_laplace_discrete} and
  \item any operations needed in order to solve the linear system resulting from the discrete variational problem.
\end{enumerate}
An excerpt of the \code{LaplaceModel} class implementation reads like this:
\begin{python} \label{code:laplace_model}
from ufl import TestFunction, TrialFunction, dx, grad, dot
from Bryne.models.model import GeneralFEMModel

class LaplaceModel(GeneralFEMModel):
    ...
    def setup_weak_form(self):
    """Implement the UFL weak form of the Laplace problem."""
      u = self.trial_function
      v = self.test_function

      self.weak_form = dot(grad(u), grad(v)) * dx == 0
    ...
    def solve_timestep(self):
    """Implements the actual linear system solve"""
      self.scheme.solve(target=self.u_h)
\end{python}
The actual assembly of the linear system and the linear solver logic itself is handled by the numerical backend, which in the current version of Bryne uses the dune-fem \cite{dednerPythonBindingsDUNEFEM2020a} Python API. Dune-fem, in turn, uses the Unified Form Language (UFL) \cite{alnaesUnifiedFormLanguage2014a} to parse the weak form input. 
The UFL syntax in \code{setup\_weak\_form()} is shared with FEniCS and Firedrake, which facilitates reading and writing Bryne models for users familiar with these libraries. 
We build on this example class to elaborate on the benefits of this approach over a monolithic Python script in the following sections. 

As shown in Fig. \ref{fig:coupling_inheritance}, our \code{LaplaceModel} is a subclass of \code{GeneralFEMModel}, which in turn derives from the abstract \code{Model} class. 
Deriving from \code{GeneralFEMModel}, we inherit attributes and methods shared by finite element solvers implemented with dune-fem. Examples include the discrete test and trial function definitions, the discrete approximation space, and a common handling of user-input boundary conditions.
All models implemented in Bryne, including \code{GeneralFEMModel}s, derive from an abstract \code{Model} class, which defines a common interface for the variables or fields a model computes and the inputs it requires to perform the computation. 
This design choice allows us to
\begin{enumerate}
  \item reuse a model in different simulation setups,
  \item combine it with other models and
  \item to extend the model through inheritance,
\end{enumerate}
once first implemented.

\subsubsection{Model cascades through inheritance}
    The object-oriented structure of Bryne models allows us to extend PDE solvers in analogy to the mathematical model hierarchy. To solve a more complex PDE containing the terms we have already written a model class for, we can simply derive a new model class from the existing one and modify the weak form. 
    A generalization of the initial Laplace model is, for instance, given by the Poisson equation with a spatially varying source term $f(\mathbf{x})$ and Dirichlet boundary conditions,
\begin{subequations}
\begin{align}\label{eq:poisson_equation}
  -\Delta u = f(\mathbf{x}) & \quad \mathbf{x} \in \Omega & \quad ,\\
  u = g & \quad \mathbf{x} \in \partial\Omega & \quad .
\end{align}
\end{subequations}
Equation \eqref{eq:poisson_equation} contains the Laplace Eqn. \eqref{eq:laplace_equation} as the special case $f(\mathbf{x})\equiv 0$. 
To represent this in code, we build a new model class \code{PoissonModel} that inherits from \code{LaplaceModel} and modifies the weak form to include the source term $f(\mathbf{x})$:
\begin{python}
from Bryne.models.laplace_model.laplace_model import LaplaceModel
...
class PoissonModel(LaplaceModel):
    ... 
    def setup_weak_form(self):
        ...
        self.external_source = ... # some value or function
        self.weak_form = dot(grad(u), grad(v)) * dx \
          == self.external_source * v * dx
\end{python}
where we have rewritten the entire weak form instead of just modifying \code{self.weak\_form} for clarity. \\ 

Even in this most simple example, the question now becomes how and where to set the actual value of the source term \code{external\_source}.
Naively, we could set it in the \code{setup\_weak\_form()} method, but this would introduce a hard dependency on the \code{PoissonModel} class code, making it hard to reuse the model in different simulation setups. 
We could try to parse a user input for \code{external\_source}, which also sacrifices flexibility. 
What if the source term is actually physically given as the solution to another equation, possibly a PDE? 

\subsection{Combining models: coupling interfaces in Bryne}

Here we describe how we can add a coupling input to our \code{PoissonModel} class to allow other models to provide the source term $f(\mathbf{x})$ at simulation runtime. Classes derived from the Bryne \code{Model} class can implement their own coupling interface through three methods,
\begin{enumerate}
  \item \code{define\_coupling\_input} to specify the names and dimensions of the inputs that a model expects. 
  \item \code{define\_coupling\_output} to provide names and dimensions of the outputs that the model provides. Other models, in turn, can use these outputs if they have a matching coupling input.
  \item \code{link\_coupling\_input} to assign coupling inputs to model attributes.
\end{enumerate}
Using these methods, we can add the source term $f(\mathbf{x})$ as a coupling input to our \code{PoissonModel} class as follows:

\begin{python}
from Bryne.models.interface import CouplingInterfaceData

def define_coupling_input(self):
  self.coupling_input_data["external_source_poisson"] = \
    CouplingInterfaceData(
        coupling_data=None, ...
        data_array_dimensions=[1],
        coupling_is_optional=True)
\end{python}
Here we specify that the \code{PoissonModel} expects a coupling input named \code{external\_source\_poisson}, which is a scalar field on the discrete domain. 
By declaring the coupling optional, we allow the model to be run even if no input with the matching name \code{external\_source\_poisson} is provided. 
For the Poisson model, this is reasonable as we can use the Laplace limit $f(\mathbf{x})\equiv 0$ as a sensible default.
Similarly, we can make the solution $u_h$ to the Poisson equation available to other models by declaring it as a coupling output:
\begin{python}
def define_coupling_output(self):
  self.coupling_output_data["poisson_scalar_field"] = \
    CouplingInterfaceData(
      coupling_data=self.u_h, ... 
      data_array_dimensions=[1])
\end{python}
With this setup, we can now combine our \code{PoissonModel} with any other model that 
\begin{itemize}
  \item provides a coupling output with the key \code{external\_source\_poisson}. Such models can be used to provide the source term $f(\mathbf{x})$.
  \item uses a coupling input with the key \code{poisson\_scalar\_field}. Such models can use the solution $u_h$ of the Poisson equation as an input.
\end{itemize}
 
To show how to use generic functions defined on the grid as coupling inputs, we now assume that our source term is given by a given scalar function 
\begin{equation} \label{eq:source_term_poisson}
  f_{\text{ex},1} = f(\mathbf{x}) = 
  \begin{cases}
    f_{\text{source}} & \text{if } \|\mathbf{x} - (0.25, 0.25)^T\|_2 < r, \\
    f_{\text{source}} & \text{if } \|\mathbf{x} - (0.75, 0.75)^T\|_2 < r, \\
    0 & \text{else,} 
  \end{cases}
\end{equation}
defined on a unit square domain $\Omega = [0, 1]^2$ with $r=0.25$. 
This spatially varying source applies a constant value of $f_{\text{source}}$ on two circular regions centered around $(0.25, 0.25)^T$ and $(0.75, 0.75)^T$. 
For functions to be analytically evaluated on the grid, Bryne provides the \code{ExternalField} class, which can provide arbitrary scalar, vector, or tensor fields on the grid with a coupling name defined by the user. 
For this to work, users have to provide a dune-fem \code{gridFunction} that takes standard UFL expressions, which allows for complex space and time-dependent function definitions. 
The implementation of the piecewise constant source term from Eqn. \eqref{eq:source_term_poisson} can be found in the \code{examples/3\_poisson\_source\_term} folder of the data repository, where the name of the field computed by \code{ExternalField} is defined in the \code{source\_model/settings.yaml} file 
\begin{lstlisting}[language=yaml]
--- !ExternalFieldSettings
field_output_name: external_source_poisson
range: 1
\end{lstlisting}
to match the expected input name that we defined for the \code{PoissonModel} class. \\

\begin{figure}
  \centering
  \includegraphics[width=\textwidth]{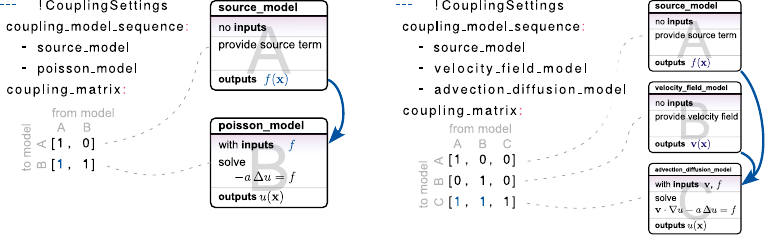}
  \caption[Model coupling setup examples]{Model coupling setup with Bryne. 
  In each timestep, models are solved in the order defined by the coupling sequence (\textit{model boxes top to bottom}). 
  If two models have a matching coupling interface, outputs of one model can be used as inputs for another model in the coupling sequence (\textit{arrows}). 
  These connections are set up via a coupling matrix, where each off-diagonal entry of one in row $i$ and column $j$ indicates that the $i$-th model in the sequence takes an input from the $j$-th model. 
  In the left example, the Poisson model \textbf{B}, second in the coupling sequence, takes the output of the external field providing the source term \textbf{A}, which comes first in the sequence. Subsequently, the entry $\left\{2,1\right\}$ of the coupling matrix is one. 
  }
  \label{fig:coupling_examples}
\end{figure}

With this setup, users can combine the two models in a model sequence. 
Model coupling is defined in \code{coupling.yaml} configuration file as sketched in Fig. \ref{fig:coupling_examples} (\textit{left}).
The coupling model sequence defines the order in which models are solved. 
Here, the first model in the sequence provides the \code{ExternalField} source term to the second \code{PoissonModel}.
In the default setting of a \textit{weak coupling}, the sequence (\textit{second column of Fig. \ref{fig:coupling_examples}}) runs once for each timestep of the simulation, exchanging data according to the connections defined using a coupling matrix. \\

\begin{figure}
  \centering
  \includegraphics[width=\textwidth]{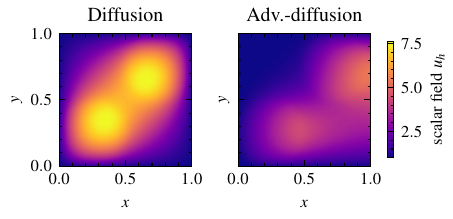}
  \caption[Model building 1: Poisson and steady advection diffusion]{Results for the first model building examples. 
  The left figure shows the numerical steady state solution for the Poisson problem with constant Dirichlet condition $g=1$ and a spatially varying source term $f(\mathbf{x})$ given by two circles of radius $r=0.25$ centered at $(x,\,y)=(0.25, 0.25)$ and $(0.75, 0.75)$ with $f_{\text{source}}=100$ (\textit{see Eqn. \eqref{eq:source_term_poisson}}). 
  On the right, a linear advection term with a constant velocity field $\mathbf{v}=(1, 0)$ is added, and a homogeneous Neumann condition is applied on the right boundary ($x=1$). 
  As a result, the solution is skewed in the direction of the external flow field. }
  \label{fig:model_examples_1_steady}
\end{figure}

In the coupling matrix, each row $i$ represents the inputs to the $i$-th model in the coupling sequence, and an input is taken from the $j$-th model in column $j$ if the entry $\left\{i,\,j\right\}$ is one. 
The diagonal entries are set to one by convention. 
In our minimal example, the coupling matrix looks like this: 
\begin{equation}
  \bordermatrix{ & \textit{from } \text{ExternalField} & \textit{from } \text{PoissonModel} \cr
      \textit{into} & \text{ExternalField} & \text{ExternalField}  \cr
      \textit{into} & \text{PoissonModel} & \text{PoissonModel} }
  = 
  \begin{bmatrix}
  1 & 0 \\
  1 & 1
\end{bmatrix} \quad .
\end{equation}
According to the coupling sequence shown in Fig. \ref{fig:coupling_examples} (\textit{left}), the first row lists inputs into the \code{ExternalField} model while the second row lists inputs into the \code{PoissonModel}.
Other than the diagonal entries, the entry at the second row and first column is one.
With this, the Bryne driver can connect the output of the \code{ExternalField} model, given by the source term $f(\mathbf{x})$, to the corresponding input of the \code{PoissonModel}. 
Ultimately, the \code{PoissonModel} can assign the input value to its own member variable \code{external\_source} in the \code{link\_coupling\_input()} method: 
\begin{python}
def link_coupling_input(self):
  ... # check if coupling input exists, then
  self.external_source = self.coupling_input_data[
      "external_source_poisson"
  ].coupling_data
\end{python}
An example solution to the Poisson model for the first example setup with constant Dirichlet boundary conditions $g=1$ and the source term from Eqn. \eqref{eq:source_term_poisson} is shown in Fig. \ref{fig:model_examples_1_steady} (\textit{left}). \\ 

To summarize, the Bryne coupling interface handles data exchange between classes derived from the Bryne \code{Model} class.
Since the term "coupling" is quite overloaded, it should be clearly stated that we do not refer to an exchange of data across interfaces or between different grids or even software packages here. 
Instead, we use "coupling" in an operator-split sense, referring to the passing of data between models defined on the same discrete grid at each timestep. 

\paragraph{Weak and strong coupling}
By default, Bryne uses weak (sometimes also called "loose") coupling, meaning that the models in the coupling sequence are solved once for each simulation timestep. 
Users can omit the coupling setup entirely to run a collection of models without any data exchange. 
In that case, models will just be run in alphabetical order of appearance, as determined by the model input folder names. 

Bryne also supports strong coupling, where the coupling sequence is iterated. 
This requires each model to define a convergence criterion based on the change in solution fields between two consecutive coupling iterations. 

\subsection{Adding model-specific parameters}
After we have combined our first two models, we show how to add parameters as user inputs through metadata files during simulation setup. After successfully adding a space-dependent source term as a generic coupling input, we would now like to add the diffusion constant $a$ to Eqn. \eqref{eq:poisson_equation}, such that our \code{PoissonModel} solves an approximation to 
\begin{equation}
  -a \, \Delta u = f(\mathbf{x}) \quad \text{in } \Omega \quad .
\end{equation}
In a first draft of a solver, users would likely just hard-code the diffusion coefficient $a$ somewhere in the \code{PoissonModel} (\textit{recall introductory figure \ref{fig:intro_notebooks}}). 
The point is now to make the input parameter $a$ traceable and modifiable without touching the model core implementation.

Adding model-specific parameters to be loaded via the Bryne simulation driver takes only a few lines of code.
The key to the parameter handling is a model-specific settings class that derives from the Bryne \code{GenericSettingsClass}. Model parameters are then simply added as class attributes in the constructor. For example, to add a constant diffusion constant, we can define 
\begin{python}
from Bryne. ... .generic_settings import GenericSettingsClass
class PoissonSettings(GenericSettingsClass):
    yaml_tag = "!PoissonSettings"

    def __init__(self, **kwargs):
        super().__init__(**kwargs)
        self.diffusion_coefficient = 1.0
\end{python}
Here, a default value of $1.0$ can be specified. Any parameter docstring will be available in the API documentation. Then, in the \code{PoissonModel} class, we only add 
\begin{python}
from Bryne.models.poisson_model.poisson_settings import PoissonSettings
class PoissonModel(Model):
    def __init__(self, **kwargs):
        ...
        self.model_settings = PoissonSettings() 
\end{python}
to make the diffusion coefficient available everywhere in the model code. To set the diffusion coefficient in our input simulation setup, we can add a \code{settings.yaml} file to the Poisson model setup folder, where users can set and check the value of $a$: 
\begin{lstlisting}[language=yaml]
--- !PoissonSettings
diffusion_coefficient: 0.5
\end{lstlisting}
If the file is not provided, $a$ will default to the value defined in the \code{PoissonSettings} class. 
At the start of each simulation, the \code{settings.yaml} file will be copied to the run output directory along with the rest of the simulation inputs, allowing users to trace the exact value of $a$ used in each individual simulation run. 

\subsection{Adding model complexity}
\textit{Full example code can be found in \code{advection\_diffusion\_example\_model.py}}, located in \code{bryne/models/advection\_diffusion\_model}. \\

With these basic building blocks in place, we can continue down the model cascade. We are reusing our existing Poisson model to build a solver for the steady linear advection-diffusion problem
\begin{subequations}\label{eq:steady_advection_diffusion}
\begin{align}
  \boldsymbol{\nabla}\cdot \left(\mathbf{b}\left(\mathbf{x}\right)\,u \right)-a \, \Delta u = f(\mathbf{x}) & \quad \mathbf{x} \in \Omega & \quad , \label{eq:steady_advection_diffusion_pde}\\
  u = g & \quad \mathbf{x} \in \Gamma_E & \quad ,\\
  \boldsymbol{\nabla}u\cdot \mathbf{n} = 0 & \quad \mathbf{x} \in \Gamma_N & \quad ,
\end{align}
\end{subequations}
where the boundary $\partial \Omega = \Gamma_E \cup \Gamma_N$ consists of disjoint Dirichlet boundary $\Gamma_E$ and natural Neumann boundary $\Gamma_N$ parts. Since the Poisson model is fully contained in the advection-diffusion equation, we can build a new \code{SteadyLinearAdvectionDiffusionModel} by deriving from the \code{PoissonModel} (\textit{recall the overview in Fig. \ref{fig:coupling_inheritance}}). \\

To use our model in combination with other models that provide a velocity field $\mathbf{b}$ as coupling output, we can add a vector of dimension $d$ with key \code{transport\_velocity} to the coupling interface in \code{define\_coupling\_input}. 
This way, we have again avoided defining the velocity field in the model code. 
Instead, we can now flexibly exchange the actual source of the velocity field. 
In a first trial, we might want to use a known velocity field and, similar to the previous example, we can configure an \code{ExternalField} to provide an analytical expression on the grid.
The corresponding coupling setup is sketched on the right of  Fig. \ref{fig:coupling_examples}, where the coupling sequence of the three models to provide the source term $f(\mathbf{x})$, the velocity field $\mathbf{b}(\mathbf{x})$ and the numerical solution of Eqn. \eqref{eq:steady_advection_diffusion} is shown.    
An example where the previous Poisson setup with a spatially varying source term is complemented by a constant advective transport with velocity $\mathbf{b}=(1, 0)^T$ is shown in Fig. \ref{fig:model_examples_1_steady} (\textit{right}). 
After testing the model with a known velocity field, we could then easily switch to a physics-based velocity field, as computed, for example, by an incompressible Navier-Stokes model.

To achieve this, the actual implementation of the model then only has to add the weak form advection term and the definition of a coupling input for the velocity field $\mathbf{b}$. 
By inheriting from an existing model, we don't have to rewrite any of the solver logic and model setup, reducing the entire implementation of the advection-diffusion model to only a few lines of effective code. 

\subsection{Transient models}
\textit{Full example code can be found in \code{transient\_advection\_diffusion\_model.py}}, located in \code{bryne/models/advection\_diffusion\_model}. \\

With the steady advection-diffusion model in place, we can now add a time dependency to Eqn. \eqref{eq:steady_advection_diffusion_pde} to obtain the transient advection-diffusion equation
\begin{equation}\label{eq:transient_advection_diffusion}
\frac{\partial u}{\partial t} + 
  \boldsymbol{\nabla}\cdot \left(\mathbf{b}\left(\mathbf{x}\right)\,u \right)-a \, \Delta u = f(\mathbf{x}) \quad \mathbf{x} \in \Omega  \quad . 
\end{equation}
The time discretization is a property of the model implementation and can be modified directly in \code{setup\_weak\_form()} by modifying the weak form according to a chosen time integration scheme.
To compute the unknown numerical solution $u_{n+1}$ at discrete time $t_{n+1}$, a simple implicit Euler time discretization
\begin{equation}\label{eq:implicit_euler}
\frac{\partial u}{\partial t} \approx \frac{u_{n+1} - u_n}{\Delta t} \quad \text{with } u_n = u(t_n) \text{ at time } t_n
\end{equation}
for a given solution $u_n$ at the previous time step $t_n$ can be implemented in UFL syntax  
\begin{python}
u = self.trial_function
v = self.test_function
time_discretization = dot((u-self.u_n)/ self.dt, v)*dx
      
self.weak_form = (
    time_discretization + a * dot(grad(u), grad(v)) * dx + ...
)
\end{python}
Parameters like the time discretization step size $\Delta t$ can be controlled per model or globally through a \code{time.yaml} settings file in the simulation setup. 
Other time discretizations are possible by implementing the corresponding weak form terms. 
Modifying the stencil from Eqn. \eqref{eq:implicit_euler}, we could, for example, easily switch the Euler scheme for a third-order BDF2 method. \\ 

Again, the actual reordering of the weak form and the linear system assembly is handled by the dune-fem backend and is not part of Bryne. To this end, users are left with the task (and freedom) to implement the actual time update by overwriting the \code{solve\_timestep()} method in the model class. 
In our example, this is a simple update along the lines of 
\begin{python}
def solve_timestep(self):
  self.time.value += self.dt.value
  self.u_n.assign(self.u_h)
  self.scheme.solve(target=self.u_h) # actual solve for u_{n+1}
  self.timestep += 1 
\end{python}
but more complex update schemes can be implemented as required by the model. 
In particular, users can write their own nonlinear solvers or non-trivial predictor-corrector schemes. 
For interested readers, an example for this is implemented in the \code{ThermochemistryDGModel} class, which is described later and distributed as part of the supplementary code. \\    

\begin{figure}
  \centering
  \includegraphics[width=\textwidth]{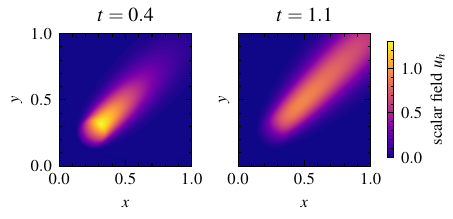}
  \caption[Model building 2: transient advection-diffusion]{The second basic example adds a time discretization to the steady advection-diffusion model and solves the model for $a=0.01$ with a spatio-temporally varying source term $f(\mathbf{x}, t)$. The source consists of the bottom-left circular source term of the steady example problem with $f_{\text{source}}=10$, which is switched off for $t>1.0$. }
  \label{fig:model_examples_1_transient}
\end{figure}

The modular design of our FEM solvers now again reduced the needed implementation effort to the actual differences between the steady and the transient versions of the advection-diffusion model. 
Since the coupling interface has been inherited, the transient solver can be directly combined with other models in the same way as its steady counterpart, for example, a coupling sequence similar to the one shown in Fig. \ref{fig:coupling_examples} (\textit{right}). 
With that, we can freely modify the source term and velocity field, both of which are part of the user simulation setup and can be arbitrary UFL functions of both space and time.
Figure \ref{fig:model_examples_1_transient} shows a demo, where the source term is modified to be a spatio-temporally varying function,
\begin{equation} \label{eq:source_term_adv_diff}
  f(\mathbf{x}) = 
  \begin{cases}
    f_{\text{source}} & \text{if } \|\mathbf{x} - (0.25, 0.25)^T\|_2 < r, \\
    0 & \text{else,} 
  \end{cases} \quad \text{for }t<1.0 \quad \text{and} \quad 0 \quad \text{for }t \geq 1.0 \quad ,
\end{equation}
build from the lower left circular source from the steady examples with $f_{\text{source}}=10$ and $r=0.25$, which is switched off for $t>1.0$. The velocity field is $\mathbf{b}=(1, 1)^T$, $a=0.01$,  and all boundaries are homogeneous Neumann boundaries, leading to the comet-trail-like transport in the downwind direction of the circular source. 

\subsection{Boundary and initial conditions}
In the previous example, we have sketched the architecture of model building in Bryne.
The question remains where and how to set the boundary conditions to close PDE problems.
Because the assembly of the linear system is handled by dune-fem, the enforcement of essential boundary conditions is closely linked to the backend library syntax.

For maximum flexibility, boundary conditions can be defined in a \code{boundary.py} file, which is part of the simulation inputs for each FEM model.
In this file, users are provided with a template function \code{setup\_boundary\_conditions()} that can be modified to set up the boundary conditions for the model. 
For example, a basic Dirichlet boundary condition for our transported scalar $u$ with a value of $1$ on the left boundary of a 2D rectangular grid could be set like this:
\begin{python}
from ufl import eq
from dune.ufl import DirichletBC, BoundaryId

def setup_boundary(model: GeneralFEMModel, boundary: ModelBoundary):
    ...
    space = model.discrete_space
    is_left = eq(BoundaryId(space), 1)
    boundary.dirichlet_boundary_eqns = \ 
            [DirichletBC(space, 1.0, is_left)]
    ...
\end{python}
At runtime, the Bryne driver will call this function, passing in each FEM model instance and its respective boundary attribute. 
Generally, boundary conditions can be passed in directly as dune-fem \code{DirichletBC} objects or by directly passing UFL values to the degrees of freedom in the solution space of the model. 
The actual processing of the boundary values, such as interpolation of the prescribed initial condition, happens in a method \code{set\_boundaries()} that all \code{GeneralFEMModel}s have to implement.
Different types of boundary conditions are managed through the \code{ModelBoundary} class.
For interested readers, we refer to the class API documentation for a more in-depth discussion. 
A variety of \code{boundary.py} examples are distributed with the attached repo, both in the \code{examples} directory and \code{tests} subfolders. 

Note that for some boundary conditions, model implementations need to handle their own processing of the information stored in the \code{ModelBoundary} class.
This includes non-homogeneous flux boundary conditions or weakly enforced Dirichlet values, the latter being common in discontinuous Galerkin (DG) methods. 
We forego a discussion here to keep this description concise, but examples can be found in the \code{StokesBrinkmanDGModel} and \code{ThermochemistryDGModel} class implementations. 

\subsection{Extending the examples}
The previous examples, summarized in Fig. \ref{fig:coupling_inheritance}, have shown the essential ingredients of Bryne FEM models. 
The example models are kept minimal and therefore lack any stabilization terms.
The toy examples nevertheless open up a range of extension possibilities, and we invite readers to try the supplementary Docker container and play with the examples.
Some ideas that could be explored:
\begin{enumerate}
  \item build your own space and time-dependent source term,
  \item make the velocity field space and time dependent (\textit{an example with a vortex-like flow field can be found in the \code{examples} directory}),
  \item extend the transient advection-diffusion model, for example, by adding a reactive term
  \item play with the time discretization to try out different time integration schemes, or
  \item modify the weak form to apply stabilization or a DG scheme to stabilize advection (\textit{a mixed DG solver for incompressible flows is implemented in the \code{StokesBrinkmanDGModel} class}).
\end{enumerate}
Below, we summarize some benefits of writing a model in Bryne over singular notebooks or scripts:
\begin{itemize}
  \item There is no need to rewrite or copy-paste any parts of the solver logic other than the call to the dune-fem (non-)linear solver. The Bryne driver provides the outer time loop, and just by adding your model to the simulation setup, it will be solved at each timestep. 
  \item Any working base version of a model can be extended by inheritance. 
  \item For coupled problems that can be solved in an operator-split fashion, models can be built and tested individually and then combined. 
  \item By using the Bryne \code{ExternalField} model to provide known fields like source terms, experimenting becomes easier since the exact definition of the field will be stored with each simulation run. 
  \item Similarly, model parameters can be exposed as user inputs to make setups easily modifiable and reproducible. 
\end{itemize}

\subsection{Simulation setup structure}

\begin{figure}
  \centering
  \includegraphics[width=0.8\textwidth]{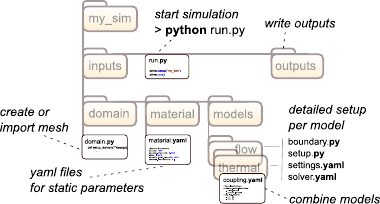}
  \caption[Structure of Bryne simulation inputs]{Structure of Bryne simulation inputs. A minimal simulation folder consists of a directory \code{inputs} providing the simulation setup and a Python script \code{run.py} that calls the Bryne simulation driver. Simulation setups are structured by simulation components, the actual input definition is realized through a combination of editable YAML files and Python scripts imported at simulation runtime.}
  \label{fig:input_structure}
\end{figure}

As shown in Fig. \ref{fig:input_structure}, Bryne simulation setups are folders with a tree structure that mirrors the building blocks of a simulation.
At the core of a Bryne simulation lies an \code{inputs} directory, which contains the input parameters grouped by simulation components. \\

Basic inputs are provided in two ways. Static inputs such as settings or parameters with simple numerical or string values can be provided in  human-readable YAML files:
\begin{lstlisting}[language=yaml]
--- !MaterialParameters
phase: liquid
density: 1000
dynamic_viscosity: 0.0013
specific_heat_capacity: 4200
thermal_conductivity: 0.6
\end{lstlisting}
Dataclasses, such as \code{MaterialParameters}, derive from \code{YAMLObject} and therefore can be instantiated by loading a YAML file with the corresponding class tag, in this case \code{!MaterialParameters}.
Documentation and default values for parameters can be provided in the class definition. For example, we can check the \code{MaterialParameters} class to find
\begin{python}
class MaterialParameters(GenericSettingsClass):
  yaml_tag = "!MaterialParameters"

  def __init__(self, **kwargs):
      super().__init__(**kwargs)
      self.thermal_conductivity = 0.6 
      """Thermal conductivity, default unit is [W / (m*K)]"""
      self.latent_heat = 333700  
      """Latent heat (of solidification), default unit is [J/kg]"""
\end{python}
The setup is robust in the sense that default values will be available at runtime even if the user does not provide a value in the YAML file. 
An advantage is that the parameter explanation in docstrings is available to the user as part of the automated API documentation. 
Examples of Bryne pre-implemented parameter classes can be found in \code{Bryne.io\_manager.settings}. 
In a later example, we will show that it is easy to add new custom parameter classes when developing new models. \\

Some parts of the simulation setup can be more complex or require Python code to be executed at runtime. One example is the setup of the computational domain, where a YAML-based approach quickly becomes limiting as it is not flexible enough to cover options for all mesh types or external mesh file imports. 
For such cases, inputs are provided as small pre-defined Python functions that are imported at runtime. For example, the \code{domain} folder contains a file \code{domain.py} with a user-modifiable function code \code{setup\_domain()} to set up the grid:

\begin{python}
from dune.grid import cartesianDomain
from dune.alugrid import aluCubeGrid as leafGridView
from Bryne.config.config import simulation_config as config

def setup_domain(**kwargs):
    # define x_min, x_max, y_min, y_max, n_x, n_y.
    # ... then
    domain = cartesianDomain([x_min, y_min],[x_max, y_max],[n_x, n_y])
    config.gridView = adaptiveGridView(leafGridView(domain))
\end{python}
We decided on this approach as it allows for maximum flexibility. For example, due to the modular design of the dune-fem backend \cite{dednerPythonBindingsDUNEFEM2020a}, changing the grid type from a quad to a triangular grid is as simple as exchanging \code{aluCubeGrid} with \code{aluConformGrid} in the import statement. 
We provide several setup examples that demonstrate usage without requiring a deep understanding of the dune-fem Python API.  \\

Even a minimial FEM simulation setup will always require a discrete domain and at least one model to solve. Other than the \code{domain}, the second essential simulation building block is the \code{models} directory. 
It contains a subfolder with the detailed parameter setup for each model to be solved in the simulation. 
The execution order and data exchange between models is configured in a \code{coupling.yaml} file, discussed later. \\

The Bryne simulation driver is created and executed in a Python script \code{run.py} located in the simulation root directory, 
\begin{python}
from Bryne.driver.driver import SimulationDriver

driver = SimulationDriver()
driver.setup()
driver.run()
\end{python}
and the simulation can then be started by executing \code{python run.py} or \code{mpirun -np 4 python run.py}. 
For every simulation run, Bryne will store a copy of the simulation \code{inputs} directory in the output directory of the simulation run. 
That way, structured metadata on the exact simulation setup is always kept with the simulation results. 
Hard-to-document simulation details such as the exact polynomial type, mesh, convergence tolerances, relaxation factors, and material parameters are hence always sustainably archived. 
Because the \code{inputs} folder archived for a given simulation contains the full simulation setup, it can be rerun as a Bryne simulation. 
This constitutes a strength of our approach, as it allows for easy reproduction given a fixed Bryne and dune-fem version. 
To pin the exact model version used to achieve a result, git metadata such as branch names and commit hashes are stored with the outputs. \\

In conjunction, our approach results in a collection of re-executable metadata for simulation setups.


\subsection{Adaptive mesh refinement}

We conclude the chapter on Bryne features with a brief discussion of mesh adaptivity. 
Figure \ref{fig:adaptivity} shows a sketch of a Bryne timestep solve for a weakly coupled model sequence.
Some dune-fem grids, such as \code{ALUGrid} \cite{alugrid:16}, support adaptive mesh refinement (AMR). Bryne builds on this to allow for $h-$refinement and coarsening. 
Currently, two types of refinement are supported:

\begin{figure}
  \centering
  \includegraphics[width=\textwidth]{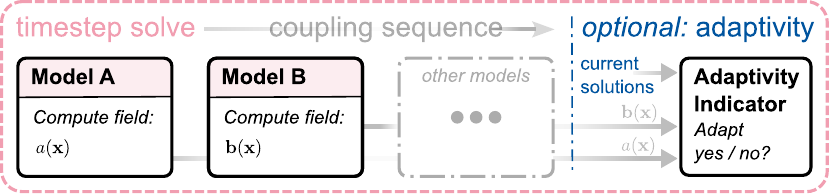}
  \caption[Sketch of a timestep solve in Bryne]{Sketch of a timestep solve in Bryne. At each timestep, a sequence of models is solved (\textit{left to right}). 
  Each model computes its solution field(s), optionally passing data to other models down the coupling sequence.
  Optionally, Bryne supports $h-$refinement for meshes that support adaptivity, such as grids based on \code{ALUGrid} \cite{alugrid:16}. 
  Users can provide custom refinement criteria, based on an indicator function that can use any current solution fields. 
  }
  \label{fig:adaptivity}
\end{figure}

\begin{itemize}
  \item \textbf{Initial refinement} criteria are evaluated once at the start of the simulation. 
  This can be used to refine around regions of interest, based on the initial condition or geometric features.
  \item \textbf{Timestep refinement} criteria are evaluated at the end of each timestep. As shown in Fig. \ref{fig:adaptivity}, users can provide custom indicator functions of the current solution fields to determine where the mesh should be refined or coarsened.
\end{itemize}
In the current implementation, all refinement criteria require a scalar indicator function $I\left(\mathbf{x}\right)$ on the grid, which is evaluated at the corresponding adaptivity call.
Bryne then passes this function and user-defined coarsening and refinement thresholds to the dune-fem backend, which will then adapt the mesh according to
\begin{equation}
  \text{do}\quad \begin{cases}
    \text{refine} & \text{if } I\left(\mathbf{x}\right) > \text{refinement threshold} \\
    \text{coarsen} & \text{if } I\left(\mathbf{x}\right) < \text{coarsening threshold} \\
    \text{nothing} & \text{otherwise}
  \end{cases} \quad .
\end{equation}
The function $I\left(\mathbf{x}\right)$ can be any space- and time-dependent UFL expression. 
In particular, it can depend on any of the current solution fields computed in the model sequence. 
The indicator function as well as coarsening and refinement thresholds, number of refinement calls (\textit{how often to refine per adaptivity call}), and maximum refinement levels can optionally be set in a subfolder \code{adaptivity} of the simulation input directory.
In the following discussion of the application examples, we show an illustrative example of a mesh refinement setup for the 1D Stefan problem.

Note that the dune-fem backend supports $p-$refinement of the finite element degree. We have not yet experimented with this feature in Bryne and forego a discussion here.

\section{Application examples}\label{sec:application_examples}
Starting from a toy PDE example, we have covered basic ingredients of the Bryne workflow.
In the following, we show some more complex examples to demonstrate how we use Bryne in our research. 
We start with a brief overview of the pre-implemented models that ship with the open-source release of Bryne and then discuss how they can be combined to build complex simulations incrementally.

\subsection{Pre-implemented models}
\textit{Python source code for the pre-implemented models can be found in \code{bryne/models/specific\_model\_name} directory of the Bryne repository.} \\

\begin{table}[h]
    \centering
    \begin{tabular}{llllll} 
        & \textit{Name} & \textit{Equation} & \multicolumn{2}{l}{\textit{Interface}} & \textit{Ref.} \\[0.1cm] 
        \hline \\[-0.1cm]
        & & & \textit{In} & \textit{Out} & \\[0.1cm] 
        \multirow{3}{*}{%
        \rotatebox[origin=c]{90}{\textit{Demo}}} & Laplace & \eqref{eq:laplace_equation} &  & $u$ &  \\ 
         & Poisson & \eqref{eq:poisson_equation} &  $f$ &  $u$ & \\ 
         & AdvectionDiffusionExample & \eqref{eq:steady_advection_diffusion} & $f$, $\mathbf{b}$ & $u$ &  \\ 
         & TransientAdvectionDiffusion & \eqref{eq:transient_advection_diffusion} &  $f$, $\mathbf{b}$ & $u(t)$ & \\[0.5cm]
        \multirow{3}{*}{%
        \rotatebox[origin=c]{90}{\textit{Research}}} & ExternalField & custom algebraic equations & user def. & user def. &  \\ 
         & Permeability & porosity-permeability relation & $\Phi$ & $\mathbf{K}(\Phi)$ & \cite{lebarsInterfacialConditionsPure2006} \\ 
        & StokesBrinkmanDG& Darcy-Brinkman-Stokes model &$ \underline{\Phi}$, $\underline{\mathbf{K}}$, $T$	& $\mathbf{u}$, $p$	& \cite{terschanskiStableRegimesMixed2025}	\\
        & ThermochemistryDG	&	solid-liquid energy conservation &$\mathbf{u}$	& $\Phi$, $T$	& \cite{kaaksEnergyconservativeDGFEMApproach2023}	\\
    \end{tabular}
    \caption{Overview of FEM models that are part of the Bryne release described in this article. 
    Models from the example section are shown first, the last four models are part of our application examples.
    The two \textit{interface} columns indicate coupling inputs and outputs of the model, as defined in the \code{define\_coupling\_input()} method of the model class. 
    Underlined variables are mandatory inputs needed to run the models. Other inputs are optional and the respective model will provide sensible defaults (usually assuming that the corresponding contribution is zero). 
    }
    \label{tab:model_table}
\end{table}

The initial open-source release of Bryne includes a set of pre-implemented solvers that demonstrate the effectiveness and value-add of the Bryne framework. 
Table \ref{tab:model_table} shows a short overview of models, including the example model hierarchy. 
Other than that, the first release comes with two solvers:
\begin{itemize}
  \item The \code{StokesBrinkmanDGModel} class implements a mixed discontinuous Galerkin FEM solver for the Darcy-Brinkman-Stokes (DBS) described in detail in our previous work \cite{terschanskiStableRegimesMixed2025}. 
  The DBS model describes incompressible flow on heterogeneous domains, where the porosity or liquid volume fraction $\Phi$ is allowed to vary between $0$ (\textit{no flow}) over $\Phi \ll 1$ (\textit{Darcy flow}) to $1$ (\textit{Navier-Stokes flow}).
  The \code{StokesBrinkmanDGModel} solver seeks approximation to the velocity field $\udarcy$ and pressure $\pdarcy$ described by the equation \cite{lebarsInterfacialConditionsPure2006}:
  \begin{subequations}\label{eq:dbs_model_strong}
  \begin{align}\label{eq:dbs_model_momentum}
  \rho_l\,\left(\frac{\partial \udarcy}{\partial t} + \udarcy\cdot\boldsymbol{\nabla}\left(\frac{\udarcy}{\Phi}\right) \right) 
  - \eta_l\,\boldsymbol{\nabla}^2\udarcy + \boldsymbol{\nabla}\pdarcy + \Phi\,\eta_l\,\mathbf{K}(\Phi)^{-1}\,\udarcy= \mathbf{f}  
  & \text{ on }\Omega^{\text{DBS}} \times (0,T]\, , \\
  \boldsymbol{\nabla}\cdot\udarcy = g  
  & \text{ on }\Omega^{\text{DBS}} \times (0,T]\, , \\
  \udarcy = \udarcy_0\,, \quad \pdarcy = p_0 
  & \text{ at } t = 0 \, .
\end{align}
\end{subequations}
Here, $\Phi$ is the porosity or liquid volume fraction field, $\mathbf{K}(\Phi)$ is the permeability tensor, and $\rho_l$, $\eta_l$ are the density and dynamic viscosity of the fluid respectively.
A core property of this model is that the Navier-Stokes equations and Darcy's law are recovered from \label{eqdbs_model_strong} in the limit $\Phi \to 1$, $\left\lVert\mathbf{K}^{-1}\right\rVert\to 0$ and $\Phi \to 0$, $\left\lVert\mathbf{K}^{-1}\right\rVert\to \infty$, respectively.
In Bryne, the body force $\mathbf{f}$ term can be used to model thermal or chemical buoyancy through a Boussinesq approximation. 
The respective temperature field $T$ is declared a coupling input in the \code{StokesBrinkmanDGModel} class, allowing the model to take the temperature field as input from arbitrary other models. 
We will later show an example of how this can be used to couple the DBS model to a thermal energy conservation model in a phase change simulation.
For a detailed discussion, description of the discretization method, and a series of validation examples, we refer to \cite{terschanskiStableRegimesMixed2025}.

Note that the code itself was not public before, but is now available as part of the Bryne release \cite{bryne_software_repo}.

  \item The \code{ThermochemistryDGModel} class implements a DG-FEM solver for the two-phase energy conservation equation:
  \begin{subequations}\label{eq:energy_conservation_bulk}
  \begin{equation}
    \frac{\partial H}{\partial t} + \boldsymbol{\nabla}\cdot \left(\mathbf{u}\,H_l - k\left(\Phi\right) \, \boldsymbol{\nabla} T \right) = 0
    \quad \text{in } \Omega \times (0,T] \quad ,
  \end{equation}
  where $\mathbf{u}$ is the liquid velocity, $k$ is the thermal conductivity. 
The bulk enthalpy $H$ of the liquid-solid mixture is composed of solid and liquid phase enthalpies $H_l$ and $H_s$ through 
  \begin{align}
    H = \Phi\,H_l + (1-\Phi)\,H_s\,, \quad k\left(\Phi\right) = \Phi\,k_l + (1-\Phi)\,k_s& \quad , \\
    H_l = \rho_l\,c_{p,l}\, T + \rho_l\,L \,,\quad H_s = \rho_s\,c_{p,s}\, T & \quad  ,
  \end{align}
  \end{subequations}
  and $0 \leq \Phi \leq 1$ indicates the liquid volume fraction. 
  The latent heat contribution $L$ enters implicitly as part of the liquid enthalpy definition, which, together with the nonlinearity of the diffusive term, makes Eqn. \eqref{eq:energy_conservation_bulk} challenging to solve by standard means.   
  The specific discretization method and iterative solution scheme to find bulk enthalpy $H$ (or temperature $T=T(H)$) and $\Phi$ to satisfy Eqn. \eqref{eq:energy_conservation_bulk} goes back to observations from \cite{swaminathanENTHALPYMETHOD1993} and is described in detail in \cite{kaaksEnergyconservativeDGFEMApproach2023}.
  This particular update scheme, implemented in the \code{ThermochemistryDGModel}, has shown to be superior to other approaches in terms of stability and required nonlinear iterations \cite{fadenOptimumEnthalpyApproach2019}. 
  In the context of classical finite elements, similar predictor-corrector schemes have been discussed in \cite{krabbenhoftImplicitMixedEnthalpy2006}, \cite{nedjarEnthalpybasedFiniteElement2002}. 
  To our knowledge, this is the first implementation of such a DG-based enthalpy method to become open-source. 
\end{itemize}
In the following, we focus on the usage of Bryne as a simulation builder. 
We thus emphasize the reusability of tested models and put less stress on numerical details.  
In-depth numerical discussion is provided in the respective references and the full source code is available as part of the supplementary code repository.
It also contains the Bryne test suite with several regression tests that check the correctness of various subcomponents of the presented models.  

\begin{figure}
  \centering
  \includegraphics[width=\textwidth]{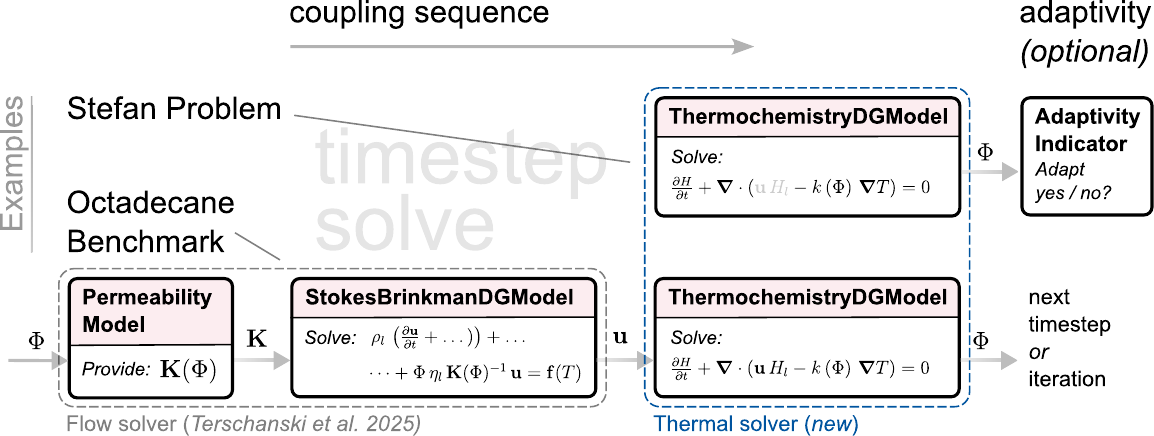}
  \caption[Results overview]{Overview of Bryne application examples presented in Section \ref{sec:convection_coupled}.
  Examples are sorted by appearance from top to bottom. 
  The coupling sequence (left to right) shows the active models and the flow of information between them. 
  In a weakly coupled simulation, the sequence is executed once per timestep solve; in an optional strongly coupled simulation, it can be iterated.
  Fields can be passed to an adaptivity indicator function for adaptive mesh refinement between timesteps.}
  \label{fig:results_overview}
\end{figure}

\subsection{Simulation building: Convection-coupled phase change}\label{sec:convection_coupled}
\textit{Simulation metadata, results and raw figures are available as part of the Zenodo data supplement} \cite{bryne_simulation_data}. \\

The final example combines the \code{StokesBrinkmanDGModel} and the \code{ThermochemistryDGModel} to build a convection-coupled phase change simulation.
The resulting simulation replicates an experimental benchmark for melting Octadecane with data from \cite{fadenOptimumEnthalpyApproach2019}.
With this, we illustrate how Bryne benefits a typical numerical modeling project, where several simulation subcomponents are combined:
\begin{itemize}
  \item the \code{StokesBrinkmanDGModel} solves the incompressible flow problem for liquid velocity $\mathbf{u}$ and pressure $p$. 
  \item a \code{Permeability} model is used to model a spatio-temporally varying permeability field $\mathbf{K}(\Phi)$ to effectively force zero flow in the solid phase ($\Phi=0$) and Stokes flow in the liquid phase ($\Phi=1$).
  \item the \code{ThermochemistryDGModel} self-consistently solves the energy conservation equation for the bulk enthalpy $H$, temperature $T$ and liquid volume fraction $\Phi$.
\end{itemize} 
The two DG-based solvers are examples of complex solvers for PDE models that include several subproblems, all of which must be tested individually. 
Development will therefore be done in several steps, with test complexity gradually increasing. 
Figure \ref{fig:results_overview} shows the sequence of models as they appear in the coupling sequence of the following examples. 
With Bryne, we can first develop and test the \code{ThermochemistryDGModel} without advection (\textit{Stefan problem, first row}). 
In a second step (\textit{Octadecane benchmark, second row}), we can combine the model with a flow solver, validated separately in \cite{terschanskiStableRegimesMixed2025}. \\

\begin{figure}
  \centering
  \includegraphics[width=0.7\textwidth]{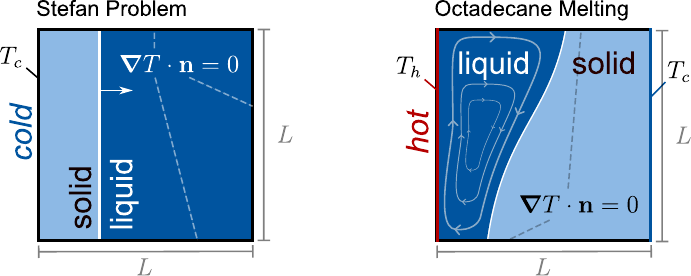}
  \caption[Application examples: setup sketches]{Setup sketches for the application examples.
  In the 1D Stefan problem (\textit{left}), a liquid, initially at constant temperature, is suddenly cooled below the freezing temperature. The scenario serves as a benchmark for solidification with purely diffusive heat transport. We build a solver for convection-coupled phase change, tested on the Octadecane benchmark (\textit{right}), by combining the thermal solver with a flow solver.
  In this setup, a solid is heated above its melting temperature at the left boundary, and natural convection causes a rotational flow in the melt.}
  \label{fig:setup_sketches}
\end{figure}

\begin{figure}
  \centering
  \includegraphics[width=\textwidth]{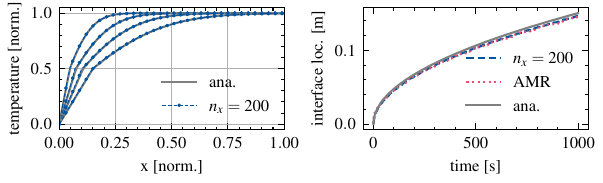}
  \caption[Stefan problem example]{Verification of the Bryne \code{ThermochemistryDGModel} on the 1D Stefan problem. The figure on the left compares the numerical temperature profile (\textit{blue dashed line}) with the analytical solution (\textit{solid gray line}) at consecutive times $t=\left\{50,\,100,\,250,\,500\right\}\,s$. 
  Kinks in the temperature profile indicate the phase change interface (PCI) location, which on the right is shown together with the analytical PCI as a function of time. 
  One numerical solution (\textit{dashed blue line}) is computed on a quadrilateral grid with $n_x \times n_y = 200 \times 1$ first-order Lagrangian finite elements and a time step size of $\Delta t = 0.5\,s$. 
  On the right, we also compare with a solution computed with adaptive mesh refinement (\textit{AMR, dotted pink line}) on a coarse base mesh of $n_x \times n_y = 25 \times 1$ quadrilateral elements, locally refined around the PCI in two bisection refinement steps. \textit{Simulation data is available in the \code{full\_stefan\_1d} and \code{full\_stefan\_1d\_with\_amr} folders in the Zenodo data supplement} \cite{bryne_simulation_data}.
  }
  \label{fig:stefan_problem}
\end{figure}

We have already discussed how Python-based discretization backends such as dune-fem offer a convenient way of adding model complexity.
Bryne adds to that a structured way of incremental model development:

\begin{enumerate}
\item In this concrete example, we started with the flow solver implemented in the \code{StokesBrinkmanDGModel}. 
We could build trust in our initial prototype method after extended testing on various 2D benchmarks \cite{terschanskiStableRegimesMixed2025}. 
With the initial incompressible solver in place, it is easy to add a linear Boussinesq buoyancy term 
\begin{equation}\label{eq:boussinesq_buoyancy}
  \mathbf{f} = \rho_l\,\Phi\,\beta\,\left(T - T_0\right)\,\mathbf{g}
\end{equation}   
as a body force term in Eqn. \eqref{eq:dbs_model_momentum}, where $\beta$ is the liquid thermal expansion coefficient, $T_0$ is a reference temperature, and $\mathbf{g}$ is the gravitational acceleration vector. 
The temperature field $T$ can then be declared as a Bryne coupling input.
\item In our case, the temperature field is to be computed from a two-phase energy conservation solver, which can be developed and tested independently.
Following \cite{kaaksEnergyconservativeDGFEMApproach2023}, the 1D Stefan problem serves as validation of the iterative solver for Eqn. \eqref{eq:energy_conservation_bulk} in the absence of advective transport ($\mathbf{u}=0$). 
Since this benchmark does not require a flow field, we can first write and test our prototype \code{ThermochemistryDGModel}, independently of a flow solver. 
\end{enumerate}
Figure \ref{fig:stefan_problem} shows the results from a validation study computed with Bryne, where the numerical solution is compared to the analytical solution of the 1D Stefan problem. 
In this setup, sketched on the left of Fig. \ref{fig:setup_sketches}, a 1D liquid rod of length $L=0.05\,m$ at initial temperature $T_0 = T(x,\,t=0) = 278\,K$ is suddenly cooled from the left boundary at $x=0$ with a constant temperature $T_c = 268\,K$ below the melting temperature $T_m = 273\,K$.
Benchmark details and reference solutions taken from \cite{kaaksEnergyconservativeDGFEMApproach2023} are repeated in the Appendix \ref{sec:stefan_setup}. 
On the left of Fig. \ref{fig:stefan_problem}, we compare the normalized numerical temperature solution $\left(T_{\text{num}} - T_c\right) / \left(T_0 - T_c\right)$ (\textit{dashed blue line}) with the corresponding analytical solution (\textit{solid gray line}). 
On the right, we show the location of the phase change interface (PCI) as a function of time, where the numerical PCI (\textit{dashed blue line}) is approximated from the $\Phi=0.5$ contour. 
At final time ($t=1000\,s$), the deviation of the numerical PCI relative to the analytical interface location is less than $1\,\%$.
An almost identical solution can be obtained with adaptive mesh refinement (AMR) (\textit{dotted pink line}), as indicated by the dotted pink line on the right of Fig. \ref{fig:stefan_problem} (\textit{for the temperature plot lines are indistinguishable visually}).
The AMR solution is computed from a coarse base mesh with $n_x \times n_y = 25 \times 1$ quadrilateral elements, implemented using dune ALUGrid \cite{alugrid:16}. The grid is locally refined and coarsened around the PCI at each timestep in two bisection steps. 
Our example refinement indicator 
\begin{equation}
  I(\mathbf{x}) = \left\lVert\boldsymbol{\nabla}\Phi\left(\mathbf{x}\right)\right\rVert_2 
  \label{eq:refinement_indicator}
\end{equation}
is based on the gradient of the liquid volume fraction $\Phi$, which is nonzero only close to the interface between the solid and liquid phases. 
We forego a detailed discussion of AMR efficiency here but include the example and simulation data to show how Bryne integrates mesh adaptivity into the simulation workflow.

\begin{enumerate}[resume]
\item To add advective transport to the energy conservation model, we can now add corresponding terms to the weak form definition in the \code{ThermochemistryDGModel} class. Again, the advection velocity is declared as an optional coupling input.
\item We can now test our DG discretization for the advective term by coupling our draft \code{ThermochemistryDGModel} to a known velocity field, similar to the introductory Example (see Fig. \ref{fig:coupling_inheritance}). At this point, the modular structure of Bryne allows us to test subcomponents of our more complex discretization without having to hard-code any velocity field in the model implementation. 
\end{enumerate}
With the flow and energy conservation models tested individually, we can now reuse our models to combine them into a single simulation setup: 
\begin{enumerate}[resume]
  \item To model flow on a domain with an implicit boundary between solid and liquid phase as indicated by the liquid volume fraction $\Phi$, we can use a permeability model with infinite permeability in the liquid phase ($\Phi=1$) and near-zero permeability in the solid phase ($\Phi=0$).  
  \item Ultimately, all models can be combined in a single coupling sequence, where the \code{StokesBrinkmanDGModel} provides the velocity field $\mathbf{u}$ to the \code{ThermochemistryDGModel} which in turn computes the temperature field $T$ entering the buoyancy term (\textit{Eqn. \eqref{eq:boussinesq_buoyancy}}) and liquid volume fraction $\Phi$ to update the permeability field $\mathbf{K}(\Phi) = k\left(\Phi\right) \mathbf{I}$ for the flow solver.
\end{enumerate}
The resulting coupling, sketched in the bottom row of Fig. \ref{fig:results_overview}, corresponds to a classical operator-split, where the hydromechanical and thermal equations are solved sequentially. Optionally, the sequence can be iterated in a strong coupling iteration until the coupling update error $\epsilon_{cp}^{\Psi} = \left\lVert\boldsymbol{\Psi}_{m+1} -  \boldsymbol{\Psi}_{m}\right\rVert_2$ of all numerical fields $\boldsymbol{\Psi}$ between the last two coupling iterations $m+1$ and $m$ is below a convergence threshold. 
By verifying that solution fields do not change significantly after the first coupling iteration, we have confirmed that the assumption of weak coupling is a reasonable approximation for the flow regime of the following benchmark setup.

\begin{table}[h]
  \centering
  \begin{tabular}{l l l l l}
    \textit{Parameter}  &  &  & \textit{Solid} & \textit{Liquid} \\[2pt]
    \midrule
    Density                       & $\rho$      & $\mathrm{kg\,m^{-3}}$ & 867      & 775.6    \\
    Specific heat capacity        & $c_p$       & $\mathrm{J\,kg^{-1}}$   & 1900     & 2240     \\
    Thermal conductivity           & $k$         & $\mathrm{W\,m^{-1}\,K^{-1}}$ & 0.32     & 0.15     \\
    Thermal expansion coefficient   &$\beta$     &$\mathrm{K^{-1}}$     & & $8.36\times 10^{-4}$   \\
    Dynamic viscosity              &$\mu$        &$\mathrm{Pa\,s}$         & & $3.75\times 10^{-3}$ \\[0.2cm] 
    Latent heat                   & $L$         & $\mathrm{J\,kg^{-1}}$      & 243680   &        \\
    Melting temperature            & $T_m$       & $\mathrm{K}$             & 301.15   &        \\

    \bottomrule
  \end{tabular}
  \caption{Material properties for the Octadecane benchmark. The same properties are used in \cite{fadenOptimumEnthalpyApproach2019}, \cite{kaaksEnergyconservativeDGFEMApproach2023}, original data is from \cite{velezTemperaturedependentThermalProperties2015} and \cite{zhangOverviewPhaseChange2010}.}
  \label{tab:octadecane_parameters}
\end{table}

\begin{figure}
  \centering
  \includegraphics[width=\textwidth]{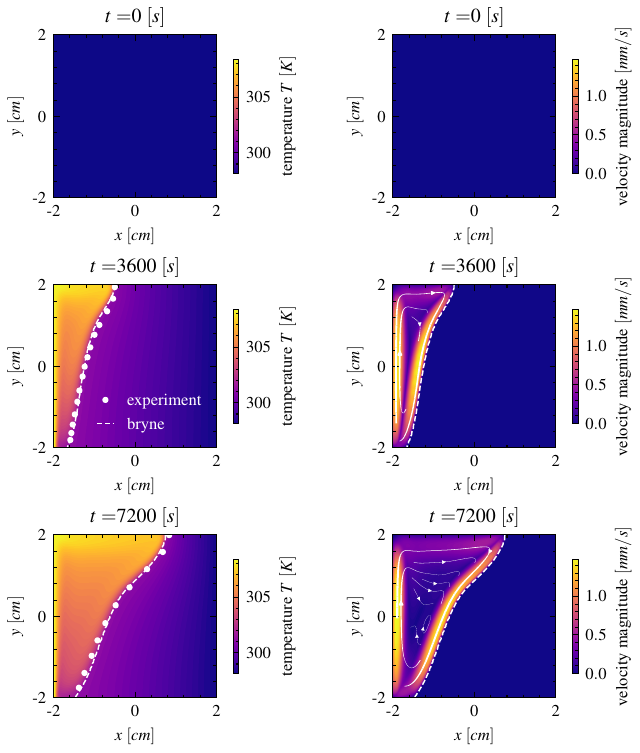}
  \caption[Octadecane benchmark: Temperature and velocity contours]{Octadecane benchmark for a convection-coupled melting simulation. The left and right columns show the temperature and velocity contours at different times $t$.
  Results are obtained on a $100\times 100$ quadrilateral grid with a nondimensional time step size corresponding to a physical time step of $\Delta t = 3.31\,s$. 
  The numerical liquid-solid phase change interface is approximated by the $\Phi = 0.5$ contour, indicated by the dashed white line. 
  White dots correspond to the experimentally measured interface location published in \cite{fadenOptimumEnthalpyApproach2019}.
  \textit{Simulation and reference data is available in the \code{octadecane} folder of the Zenodo data supplement} \cite{bryne_simulation_data}.}
  \label{fig:octadecane_melting_contours}
\end{figure}

Fig. \ref{fig:setup_sketches} (\textit{right}) shows the numerical benchmark configuration corresponding to the experimental setup from \cite{fadenOptimumEnthalpyApproach2019}.
A $\left[-0.02,\,0.02\right]^2\,m$ square domain with adiabatic top and bottom walls is filled with Octadecane, initially solid at constant temperature $T_0=298.15\,K$ (\textit{top row of Fig. \ref{fig:octadecane_melting_contours}}). While the right boundary ($x=0.02\,m$) is kept at a constant temperature $T_c=298.15\,K$, the left boundary ($x=-0.02\,m$) is suddenly heated to a constant temperature $T_h=308.15\,K$ above the melting point $T_m = 301.15\,K$.
This causes the solid to melt, and the subsequent heating of the liquid phase forces natural convection.
The resulting clockwise flow field accelerates melting at the top of the cavity, and a non-trivial liquid-solid phase change interface (PCI) develops. \\

\begin{figure}
  \centering
  \includegraphics[width=\textwidth]{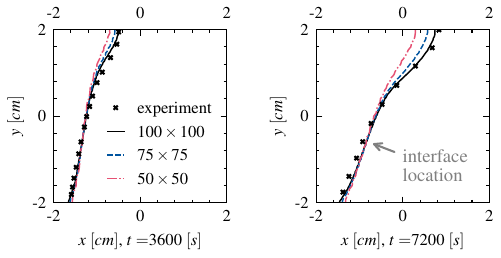}
  \caption[Octadecane benchmark: Melting interface location]{Location of the liquid-solid phase change interface at two representative time snapshots. 
  The approximate numerical interface location computed with Bryne at different mesh resolutions is recovered from the $\Phi = 0.5$ contour.
  The reference data comes from experimental measurements published in \cite{fadenOptimumEnthalpyApproach2019}.
  }
  \label{fig:octadecane_melting_pci}
\end{figure}

Fig. \ref{fig:octadecane_melting_contours} shows the temperature and velocity contours at different times $t$, obtained on a $100\times 100$ quadrilateral grid with discontinuous Lagrangian elements of order $\left\{2, 1, 1\right\}$ for velocity, pressure, and temperature, respectively. Material properties are summarized in Table \ref{tab:octadecane_parameters}.
The timestep size is chosen as $\Delta t = 0.01\,t_{\nu}$, where $t_{\nu} = \left(\rho_l \, L^2 \right) / \mu_l \approx 330.92\,s$ is the viscous timescale for the cavity with side length $L=0.04\,m$. 
The interface between solid and liquid phases is approximated by the contour of $\Phi = 0.5$, which is indicated by the dashed white line. 
White dots show the interface location as measured in \cite{fadenOptimumEnthalpyApproach2019}. 
Temperature and flow field as well as the interface location agree well with the experimental data and numerical campaigns from \cite{fadenOptimumEnthalpyApproach2019} and \cite{kaaksEnergyconservativeDGFEMApproach2023}.
For the latter, a small mesh convergence study is shown in Fig. \ref{fig:octadecane_melting_pci}, where the numerical PCI location obtained from a sequence of refined quadrilateral meshes is compared to the experimental data from \cite{fadenOptimumEnthalpyApproach2019} at two times.
In agreement with previous observations, the enthalpy-based nonlinear update scheme converges reliably in $\mathcal{O}(10)$ nonlinear iterations per timestep. 
However, the rigorous weak enforcement of boundary values for the DG scheme needed to obtain the results from Fig. \ref{fig:octadecane_melting_contours} and Fig. \ref{fig:octadecane_melting_pci} adversely affects the conditioning of the linear system, resulting in costly linear solves in each nonlinear iteration. 
In this study, we have focused on an initial physical correctness check. Improving the linear solver efficiency and scalability will be part of a future work.

\section{Conclusion}\label{sec13}
In this article, we presented the Bryne Python software framework for reusable FEM models implemented in dune-fem. 
With this, we address a gap between the rapid-prototyping capabilities of open-source FEM backends and the need for a reproducibility-enabled infrastructure to tackle complex simulation setups.
Bryne enhances the reusability of dune-fem model-based workflows by adding three dimensions of model reusability:
\begin{enumerate}
  \item wrapping core components of FEM simulations in an object-oriented architecture allows writing FEM solvers that can be reused, either in conjunction with other models or by extending the model through inheritance.
  \item Our first implementation of a model coupling interface allows building operator-split multiphysics simulations from models defined on the same grid.
  \item Simulation inputs can be provided by combining YAML files for static parameters and user-modifiable Python scripts. 
  This allows for sustainable archiving of simulations and setups that are easier to read, modify, and reproduce.
\end{enumerate}
Starting from a minimal example, we have shown what comprises a Bryne FEM model and how the object-oriented approach allows for building upon existing models incrementally. 
This approach has facilitated our research by providing efficient infrastructure for reproducible workflows. 
The software framework enables complex multiphysics models built from individually developed and tested components. 
This was demonstrated with the help of a convection-coupled phase change simulation that combines a flow and a thermal solver.
For both models, now made open-source for the first time, Bryne handles the complex simulation setup and solver logic.
This allowed us as model developers to focus on the model implementation and testing.
Using Bryne, the release of these models along with the entire setup of verification examples comes at no additional development cost. \\

In Bryne's first release, we focused on finite element models built from the dune-fem Python API. 
Many of the concepts presented here translate to other Python FEM frameworks such as FEniCS or Firedrake.
Extending the Bryne model interface to allow for FEniCS or Firedrake FEM models would be an interesting future step.
While we have only considered data exchange across models defined on the same mesh, more complex couplings could be realized through an external coupling library such as preCICE \cite{preCICEv2}.

\backmatter

\section*{Declarations}

\subsection*{Funding}
This work was funded by the German Research Foundation (DFG), Germany - grant number 460819306. 
Robert Kl\"ofkorn was supported by the Swedish Research Council through grant AI-Twin (2024-04904).

\subsection*{Data availability}
Simulation data, including inputs and paraview files, along with the raw result figures are available on Zenodo \cite{bryne_simulation_data}: \url{https://doi.org/10.5281/zenodo.15850111}

\subsection*{Code availability}
Bryne is open-source software under the GNU General Public License v3.0 or later. 
The active source repository can be found at 
\url{https://git.rwth-aachen.de/mbd/bryne} \\

This article refers to release version \code{1.0.0}, which is pinned at the following Zenodo repository \cite{bryne_software_repo}:
\url{https://doi.org/10.5281/zenodo.15789249} 

We provide a Bryne Docker container including a fixed compatible dune-fem version to facilitate reproduction.
The container can host a JupyterLab for users who just want to quickly review the examples or the source code. 
Instructions on how to use the container are included in the documentation distributed on Zenodo.

\subsection*{Acknowledgements}
We thank Ingo Steldermann for his helpful comments on the manuscript draft.

\subsection*{Author contribution}
Benjamin Terschanski: Writing - original draft, Visualization, Validation, Software, Methodology, Investigation, Formal analysis, Data curation, Conceptualization. Robert Klöfkorn: Writing - review \& editing, Software. Andreas Dedner: Writing - review
\& editing, Software. Julia Kowalski: Writing - review \& editing, Supervision, Resources, Funding acquisition, Conceptualization.

\subsection*{Use of editing tools and large language models}
The authors acknowledge the use of
Grammarly \url{https://grammarly.com} and Languagetool \url{https://languagetool.org/} for editing individual text passages.
Github Copilot \url{https://github.com/features/copilot} has been used to assist with writing Python code. 

\subsection*{Competing interests}
The authors declare no conflict of interest.

\begin{appendices}

  \section{Stefan problem benchmark}\label{sec:stefan_setup}

  Here, we repeat the setup of the Stefan problem from Section \ref{sec:application_examples} for convenience. Material parameters are summarized in Table \ref{tab:stefan_parameters}. 
  The setup sketch for this problem is shown on the left of Fig. \ref{fig:setup_sketches}.

  \begin{table}[h]
  \centering
  \begin{tabular}{l l l l l}
    \textit{Parameter}  &  &  & \textit{Solid} & \textit{Liquid} \\[2pt]
    \midrule
    Density                       & $\rho$      & $\mathrm{kg\,m^{-3}}$ & 1000      & 1000    \\
    Specific heat capacity        & $c_p$       & $\mathrm{J\,kg^{-1}}$   & 2100     & 4200     \\
    Thermal conductivity           & $k$         & $\mathrm{W\,m^{-1}\,K^{-1}}$ & 2.16     & 0.575    \\[0.2cm] 
    Latent heat                   & $L$         & $\mathrm{J\,kg^{-1}}$      & 333000   &        \\
    Melting temperature            & $T_m$       & $\mathrm{K}$             & 273.0   &        \\

    \bottomrule
  \end{tabular}
  \caption{Material properties for the 1D Stefan problem from \cite{kaaksEnergyconservativeDGFEMApproach2023}.}
  \label{tab:stefan_parameters}
\end{table}

  The analytical solution is then computed as follows  \cite{kaaksEnergyconservativeDGFEMApproach2023}, \cite{stefanProblemRubenstein}, \cite{vollerAccurateSolutionsMoving1981a}:
  At given time $t$, the interface location (\textit{vertical white line on the left of Fig. \ref{fig:setup_sketches}}) is computed from 
  \begin{equation}
    s(t) = 2\,\lambda\,\sqrt{\alpha_s\,t}
    \label{eq:stefan_pci}\quad ,
  \end{equation}
  where the similarity parameter $\lambda$ is found numerically as the root of the function
  \begin{equation}
G(\lambda) = - \frac{k_l}{k_s}\,
  \frac{
    \sqrt{\alpha_s} (T_0 - T_m) \exp\left(-\frac{\alpha_s}{\alpha_l} \lambda^2\right)
  }{
    \sqrt{\alpha_l} (T_m - T_c) \operatorname{erfc}\left(\lambda \sqrt{\frac{k_s}{k_l}}\right)
  }
+ \frac{\exp(-\lambda^2)}{\operatorname{erf}(\lambda)}
- \frac{\lambda L \sqrt{\pi}}{c_{p,s} (T_m - T_c)} \overset{!}{=}0 \, .
\end{equation}
Here $T_0 = T\left(x,t=0\right)$ is the initial temperature of the liquid phase, $T_m$ is the melting temperature, and $T_c$ is the temperature at the cooled wall boundary ($x=0$). The material parameters $k$, $L$ and $\alpha = k / \left(\rho\, c_p\right)$ are the thermal conductivity, latent heat, and thermal diffusivity respectively. 
The subscript $\bullet_s$ and $\bullet_l$ distinguishes solid and liquid phase parameters. \\

The transient analytical temperature profile can be computed for a given interface location $s(t)$ as 
\begin{equation}
T(x, t) = 
\begin{cases}
\displaystyle
\frac{T_m - T_c}{\operatorname{erf}(\lambda)}\, \operatorname{erf}\left( \frac{x}{2\sqrt{\alpha_s t}} \right) + T_c & x < s(t)\\[3ex]
\displaystyle
T_0 - 
\frac{T_0 - T_m}{\operatorname{erfc}\left( \lambda \sqrt{\frac{\alpha_s}{\alpha_l}} \right)}
\, \operatorname{erfc}\left( \frac{x}{2\sqrt{\alpha_l t}} \right) & x \geq s(t) \quad .
\end{cases}
\end{equation}
Here, we note a small typo in the corresponding analytical temperature profile given in Eqn. $(43)$ of \cite{kaaksEnergyconservativeDGFEMApproach2023}, where the root in the first denominator is printed as $\alpha_s / \sqrt{\alpha_l}$ instead of $\sqrt{\alpha_s / \alpha_l}$. 
  See the original Eqn. $(2.6)$ of \cite{vollerAccurateSolutionsMoving1981a} for reference.



\end{appendices}


\bibliography{sn-bibliography}


\begin{thebibliography}{34}
\ifx \bisbn   \undefined \def \bisbn  #1{ISBN #1}\fi
\ifx \binits  \undefined \def \binits#1{#1}\fi
\ifx \bauthor  \undefined \def \bauthor#1{#1}\fi
\ifx \batitle  \undefined \def \batitle#1{#1}\fi
\ifx \bjtitle  \undefined \def \bjtitle#1{#1}\fi
\ifx \bvolume  \undefined \def \bvolume#1{\textbf{#1}}\fi
\ifx \byear  \undefined \def \byear#1{#1}\fi
\ifx \bissue  \undefined \def \bissue#1{#1}\fi
\ifx \bfpage  \undefined \def \bfpage#1{#1}\fi
\ifx \blpage  \undefined \def \blpage #1{#1}\fi
\ifx \burl  \undefined \def \burl#1{\textsf{#1}}\fi
\ifx \doiurl  \undefined \def \doiurl#1{\url{https://doi.org/#1}}\fi
\ifx \betal  \undefined \def \betal{\textit{et al.}}\fi
\ifx \binstitute  \undefined \def \binstitute#1{#1}\fi
\ifx \binstitutionaled  \undefined \def \binstitutionaled#1{#1}\fi
\ifx \bctitle  \undefined \def \bctitle#1{#1}\fi
\ifx \beditor  \undefined \def \beditor#1{#1}\fi
\ifx \bpublisher  \undefined \def \bpublisher#1{#1}\fi
\ifx \bbtitle  \undefined \def \bbtitle#1{#1}\fi
\ifx \bedition  \undefined \def \bedition#1{#1}\fi
\ifx \bseriesno  \undefined \def \bseriesno#1{#1}\fi
\ifx \blocation  \undefined \def \blocation#1{#1}\fi
\ifx \bsertitle  \undefined \def \bsertitle#1{#1}\fi
\ifx \bsnm \undefined \def \bsnm#1{#1}\fi
\ifx \bsuffix \undefined \def \bsuffix#1{#1}\fi
\ifx \bparticle \undefined \def \bparticle#1{#1}\fi
\ifx \barticle \undefined \def \barticle#1{#1}\fi
\bibcommenthead
\ifx \bconfdate \undefined \def \bconfdate #1{#1}\fi
\ifx \botherref \undefined \def \botherref #1{#1}\fi
\ifx \url \undefined \def \url#1{\textsf{#1}}\fi
\ifx \bchapter \undefined \def \bchapter#1{#1}\fi
\ifx \bbook \undefined \def \bbook#1{#1}\fi
\ifx \bcomment \undefined \def \bcomment#1{#1}\fi
\ifx \oauthor \undefined \def \oauthor#1{#1}\fi
\ifx \citeauthoryear \undefined \def \citeauthoryear#1{#1}\fi
\ifx \endbibitem  \undefined \def \endbibitem {}\fi
\ifx \bconflocation  \undefined \def \bconflocation#1{#1}\fi
\ifx \arxivurl  \undefined \def \arxivurl#1{\textsf{#1}}\fi
\csname PreBibitemsHook\endcsname

\bibitem[\protect\citeauthoryear{Arndt et~al.}{2021}]{dealii2019design}
\begin{barticle}
\bauthor{\bsnm{Arndt}, \binits{D.}},
\bauthor{\bsnm{Bangerth}, \binits{W.}},
\bauthor{\bsnm{Davydov}, \binits{D.}},
\bauthor{\bsnm{Heister}, \binits{T.}},
\bauthor{\bsnm{Heltai}, \binits{L.}},
\bauthor{\bsnm{Kronbichler}, \binits{M.}},
\bauthor{\bsnm{Maier}, \binits{M.}},
\bauthor{\bsnm{Pelteret}, \binits{J.-P.}},
\bauthor{\bsnm{Turcksin}, \binits{B.}},
\bauthor{\bsnm{Wells}, \binits{D.}}:
\batitle{The {deal.II} finite element library: Design, features, and insights}.
\bjtitle{Computers \& Mathematics with Applications}
\bvolume{81},
\bfpage{407}--\blpage{422}
(\byear{2021})
\doiurl{10.1016/j.camwa.2020.02.022}
\end{barticle}
\endbibitem

\bibitem[\protect\citeauthoryear{Kondov
  et~al.}{2013}]{kondovMultiscaleModellingMethods2013}
\begin{bbook}
\beditor{\bsnm{Kondov}, \binits{I.}},
\beditor{\bsnm{Sutmann}, \binits{G.}},
\beditor{\bsnm{{Centre Europ{\'e}en de Calcul Atomique et Mol{\'e}culaire}}}
  (eds.):
\bbtitle{Multiscale Modelling Methods for Applications in Materials Science:
  {{CECAM}} Tutorial, 16 - 20 {{September}} 2013, {{Forschungszentrum
  J{\"u}lich}} ; Lecture Notes}.
\bsertitle{Schriften Des {{Forschungszentrums J{\"u}lich}} : [...], {{IAS}}
  Series},
vol. \bseriesno{Vol. 19}.
\bpublisher{Forschungszentrum, Zentralbibliothek},
\blocation{J{\"u}lich}
(\byear{2013})
\end{bbook}
\endbibitem

\bibitem[\protect\citeauthoryear{Baratta
  et~al.}{}]{barattaDOLFINxNextGeneration}
\begin{botherref}
\oauthor{\bsnm{Baratta}, \binits{I.A.}},
\oauthor{\bsnm{Dean}, \binits{J.P.}},
\oauthor{\bsnm{Dokken}, \binits{J.S.}},
\oauthor{\bsnm{Habera}, \binits{M.}},
\oauthor{\bsnm{Hale}, \binits{J.S.}},
\oauthor{\bsnm{Richardson}, \binits{C.N.}},
\oauthor{\bsnm{Rognes}, \binits{M.E.}},
\oauthor{\bsnm{Scroggs}, \binits{M.W.}},
\oauthor{\bsnm{Sime}, \binits{N.}},
\oauthor{\bsnm{Wells}, \binits{G.N.}}:
{{DOLFINx}}: {{The}} next generation {{FEniCS}} problem solving environment
\end{botherref}
\endbibitem

\bibitem[\protect\citeauthoryear{Rathgeber
  et~al.}{2017}]{rathgeberFiredrakeAutomatingFinite2017}
\begin{barticle}
\bauthor{\bsnm{Rathgeber}, \binits{F.}},
\bauthor{\bsnm{Ham}, \binits{D.A.}},
\bauthor{\bsnm{Mitchell}, \binits{L.}},
\bauthor{\bsnm{Lange}, \binits{M.}},
\bauthor{\bsnm{Luporini}, \binits{F.}},
\bauthor{\bsnm{Mcrae}, \binits{A.T.T.}},
\bauthor{\bsnm{Bercea}, \binits{G.-T.}},
\bauthor{\bsnm{Markall}, \binits{G.R.}},
\bauthor{\bsnm{Kelly}, \binits{P.H.J.}}:
\batitle{Firedrake: {{Automating}} the {{Finite Element Method}} by {{Composing
  Abstractions}}}.
\bjtitle{ACM Transactions on Mathematical Software}
\bvolume{43}(\bissue{3}),
\bfpage{1}--\blpage{27}
(\byear{2017})
\doiurl{10.1145/2998441}
\end{barticle}
\endbibitem

\bibitem[\protect\citeauthoryear{Bastian
  et~al.}{2021}]{bastianDuneFrameworkBasic2021}
\begin{barticle}
\bauthor{\bsnm{Bastian}, \binits{P.}},
\bauthor{\bsnm{Blatt}, \binits{M.}},
\bauthor{\bsnm{Dedner}, \binits{A.}},
\bauthor{\bsnm{Dreier}, \binits{N.-A.}},
\bauthor{\bsnm{Engwer}, \binits{C.}},
\bauthor{\bsnm{Fritze}, \binits{R.}},
\bauthor{\bsnm{Gr{\"a}ser}, \binits{C.}},
\bauthor{\bsnm{Gr{\"u}ninger}, \binits{C.}},
\bauthor{\bsnm{Kempf}, \binits{D.}},
\bauthor{\bsnm{Kl{\"o}fkorn}, \binits{R.}},
\bauthor{\bsnm{Ohlberger}, \binits{M.}},
\bauthor{\bsnm{Sander}, \binits{O.}}:
\batitle{The {{Dune}} framework: {{Basic}} concepts and recent developments}.
\bjtitle{Computers \& Mathematics with Applications}
\bvolume{81},
\bfpage{75}--\blpage{112}
(\byear{2021})
\doiurl{10.1016/j.camwa.2020.06.007}
\end{barticle}
\endbibitem

\bibitem[\protect\citeauthoryear{Dedner
  et~al.}{2020}]{dednerPythonBindingsDUNEFEM2020a}
\begin{botherref}
\oauthor{\bsnm{Dedner}, \binits{A.}},
\oauthor{\bsnm{Kloefkorn}, \binits{R.}},
\oauthor{\bsnm{Nolte}, \binits{M.}}:
Python {{Bindings}} for the {{DUNE-FEM}} Module.
Zenodo
(2020).
\doiurl{10.5281/ZENODO.3706994}
\end{botherref}
\endbibitem

\bibitem[\protect\citeauthoryear{Aln{\ae}s
  et~al.}{2014}]{alnaesUnifiedFormLanguage2014a}
\begin{barticle}
\bauthor{\bsnm{Aln{\ae}s}, \binits{M.S.}},
\bauthor{\bsnm{Logg}, \binits{A.}},
\bauthor{\bsnm{{\O}lgaard}, \binits{K.B.}},
\bauthor{\bsnm{Rognes}, \binits{M.E.}},
\bauthor{\bsnm{Wells}, \binits{G.N.}}:
\batitle{Unified form language: {{A}} domain-specific language for weak
  formulations of partial differential equations}.
\bjtitle{ACM Transactions on Mathematical Software}
\bvolume{40}(\bissue{2}),
\bfpage{1}--\blpage{37}
(\byear{2014})
\doiurl{10.1145/2566630}
\end{barticle}
\endbibitem

\bibitem[\protect\citeauthoryear{Ltd.}{2025}]{FEATool2025}
\begin{botherref}
\oauthor{\bsnm{Ltd.}, \binits{P.S.}}:
FEATool Multiphysics.
Accessed: 2025-06-17
(2025).
\url{https://www.featool.com}
\end{botherref}
\endbibitem

\bibitem[\protect\citeauthoryear{}{}]{multiphenicsx}
\begin{botherref}
Multiphenicsx.
Accessed: 2025-06-17.
\url{https://multiphenics.github.io}
\end{botherref}
\endbibitem

\bibitem[\protect\citeauthoryear{Bergersen
  et~al.}{2020}]{bergersenTurtleFSIRobustMonolithic2020}
\begin{barticle}
\bauthor{\bsnm{Bergersen}, \binits{A.}},
\bauthor{\bsnm{Slyngstad}, \binits{A.}},
\bauthor{\bsnm{Gjertsen}, \binits{S.}},
\bauthor{\bsnm{Souche}, \binits{A.}},
\bauthor{\bsnm{{Valen-Sendstad}}, \binits{K.}}:
\batitle{{{turtleFSI}}: {{A Robust}} and {{Monolithic FEniCS-based
  Fluid-Structure Interaction Solver}}}.
\bjtitle{Journal of Open Source Software}
\bvolume{5}(\bissue{50}),
\bfpage{2089}
(\byear{2020})
\doiurl{10.21105/joss.02089}
\end{barticle}
\endbibitem

\bibitem[\protect\citeauthoryear{{\"O}{\u g}{\"u}{\c c}
  et~al.}{2025}]{ogucFeVAcSPackageVisualizing2025}
\begin{barticle}
\bauthor{\bsnm{{\"O}{\u g}{\"u}{\c c}}, \binits{M.}},
\bauthor{\bsnm{Okyar}, \binits{A.F.}},
\bauthor{\bsnm{Khajah}, \binits{T.}}:
\batitle{{{FeVAcS}}: {{A}} package for visualizing acoustic scattering from
  {{1D}} periodic obstacles}.
\bjtitle{Software Impacts}
\bvolume{24},
\bfpage{100756}
(\byear{2025})
\doiurl{10.1016/j.simpa.2025.100756}
\end{barticle}
\endbibitem

\bibitem[\protect\citeauthoryear{Quintino and
  Tiago}{2025}]{quintinoOpytimalPythonFEniCS2025}
\begin{botherref}
\oauthor{\bsnm{Quintino}, \binits{N.}},
\oauthor{\bsnm{Tiago}, \binits{J.}}:
Opytimal - {{A Python}}/{{FEniCS}} Framework to Solve {{PDE-based}} Optimal
  Control Problems Considering Multiple Controls in {{2D}} and {{3D}} Domains.
In Review
(2025).
\doiurl{10.21203/rs.3.rs-6546784/v1}
\end{botherref}
\endbibitem

\bibitem[\protect\citeauthoryear{Wintzer
  et~al.}{2025}]{wintzerMultiphysicsSimulationCrystal2025}
\begin{barticle}
\bauthor{\bsnm{Wintzer}, \binits{A.}},
\bauthor{\bsnm{Abali}, \binits{B.E.}},
\bauthor{\bsnm{Dadzis}, \binits{K.}}:
\batitle{Multiphysics simulation of crystal growth with moving boundaries in
  {{FEniCS}}}.
\bjtitle{Computer Methods in Applied Mechanics and Engineering}
\bvolume{437},
\bfpage{117783}
(\byear{2025})
\doiurl{10.1016/j.cma.2025.117783}
\end{barticle}
\endbibitem

\bibitem[\protect\citeauthoryear{Kjeldsberg
  et~al.}{2023}]{kjeldsbergVerifiedValidatedMoving2023}
\begin{barticle}
\bauthor{\bsnm{Kjeldsberg}, \binits{H.A.}},
\bauthor{\bsnm{Sundnes}, \binits{J.}},
\bauthor{\bsnm{Valen-Sendstad}, \binits{K.}}:
\batitle{A verified and validated moving domain computational fluid dynamics
  solver with applications to cardiovascular flows}.
\bjtitle{International Journal for Numerical Methods in Biomedical Engineering}
\bvolume{39}(\bissue{6}),
\bfpage{3703}
(\byear{2023})
\doiurl{10.1002/cnm.3703}
\end{barticle}
\endbibitem

\bibitem[\protect\citeauthoryear{Kyas
  et~al.}{2022}]{kyasAcceleratedReactiveTransport2022}
\begin{barticle}
\bauthor{\bsnm{Kyas}, \binits{S.}},
\bauthor{\bsnm{Volpatto}, \binits{D.}},
\bauthor{\bsnm{Saar}, \binits{M.O.}},
\bauthor{\bsnm{Leal}, \binits{A.M.M.}}:
\batitle{Accelerated reactive transport simulations in heterogeneous porous
  media using {{Reaktoro}} and {{Firedrake}}}.
\bjtitle{Computational Geosciences}
\bvolume{26}(\bissue{2}),
\bfpage{295}--\blpage{327}
(\byear{2022})
\doiurl{10.1007/s10596-021-10126-2}
\end{barticle}
\endbibitem

\bibitem[\protect\citeauthoryear{Dedner
  et~al.}{2019}]{dednerPythonFrameworkHpadaptive2019a}
\begin{barticle}
\bauthor{\bsnm{Dedner}, \binits{A.}},
\bauthor{\bsnm{Kane}, \binits{B.}},
\bauthor{\bsnm{Kl{\"o}fkorn}, \binits{R.}},
\bauthor{\bsnm{Nolte}, \binits{M.}}:
\batitle{Python framework for hp-adaptive discontinuous {{Galerkin}} methods
  for two-phase flow in porous media}.
\bjtitle{Applied Mathematical Modelling}
\bvolume{67},
\bfpage{179}--\blpage{200}
(\byear{2019})
\doiurl{10.1016/j.apm.2018.10.013}
\end{barticle}
\endbibitem

\bibitem[\protect\citeauthoryear{Terschanski
  et~al.}{2025}]{terschanskiStableRegimesMixed2025}
\begin{barticle}
\bauthor{\bsnm{Terschanski}, \binits{B.}},
\bauthor{\bsnm{Kl{\"o}fkorn}, \binits{R.}},
\bauthor{\bsnm{Dedner}, \binits{A.}},
\bauthor{\bsnm{Kowalski}, \binits{J.}}:
\batitle{Stable across regimes: {{A}} mixed {{DG}} method for
  {{Darcy}}--{{Brinkman}}--{{Stokes}} type flows}.
\bjtitle{Computer Methods in Applied Mechanics and Engineering}
\bvolume{442},
\bfpage{117962}
(\byear{2025})
\doiurl{10.1016/j.cma.2025.117962}
\end{barticle}
\endbibitem

\bibitem[\protect\citeauthoryear{Dalcin et~al.}{2011}]{petsc4py:11}
\begin{barticle}
\bauthor{\bsnm{Dalcin}, \binits{L.D.}},
\bauthor{\bsnm{Paz}, \binits{R.R.}},
\bauthor{\bsnm{Kler}, \binits{P.A.}},
\bauthor{\bsnm{Cosimo}, \binits{A.}}:
\batitle{{Parallel distributed computing using Python}}.
\bjtitle{Advances in Water Resources}
\bvolume{34}(\bissue{9}),
\bfpage{1124}--\blpage{1139}
(\byear{2011})
\doiurl{10.1016/j.advwatres.2011.04.013}
\end{barticle}
\endbibitem

\bibitem[\protect\citeauthoryear{Balay et~al.}{2025}]{petsc3_22}
\begin{botherref}
\oauthor{\bsnm{Balay}, \binits{S.}}, et al.:
PETSc/TAO Users Manual,
(2025).
\doiurl{10.2172/2476320}
\end{botherref}
\endbibitem

\bibitem[\protect\citeauthoryear{Parkinson
  et~al.}{2020}]{parkinsonModellingBinaryAlloy2020}
\begin{barticle}
\bauthor{\bsnm{Parkinson}, \binits{J.R.G.}},
\bauthor{\bsnm{Martin}, \binits{D.F.}},
\bauthor{\bsnm{Wells}, \binits{A.J.}},
\bauthor{\bsnm{Katz}, \binits{R.F.}}:
\batitle{Modelling binary alloy solidification with adaptive mesh refinement}.
\bjtitle{Journal of Computational Physics: X}
\bvolume{5},
\bfpage{100043}
(\byear{2020})
\doiurl{10.1016/j.jcpx.2019.100043}
\end{barticle}
\endbibitem

\bibitem[\protect\citeauthoryear{Kaaks
  et~al.}{2023}]{kaaksEnergyconservativeDGFEMApproach2023}
\begin{barticle}
\bauthor{\bsnm{Kaaks}, \binits{B.J.}},
\bauthor{\bsnm{Rohde}, \binits{M.}},
\bauthor{\bsnm{Kloosterman}, \binits{J.-L.}},
\bauthor{\bsnm{Lathouwers}, \binits{D.}}:
\batitle{An energy-conservative {{DG-FEM}} approach for solid--liquid phase
  change}.
\bjtitle{Numerical Heat Transfer, Part B: Fundamentals}
\bvolume{0}(\bissue{0}),
\bfpage{1}--\blpage{27}
(\byear{2023})
\doiurl{10.1080/10407790.2023.2211231}
\end{barticle}
\endbibitem

\bibitem[\protect\citeauthoryear{Terschanski
  et~al.}{2025}]{bryne_software_repo}
\begin{botherref}
\oauthor{\bsnm{Terschanski}, \binits{B.}},
\oauthor{\bsnm{Klöfkorn}, \binits{R.}},
\oauthor{\bsnm{Dedner}, \binits{A.}},
\oauthor{\bsnm{Kowalski}, \binits{J.}}:
Bryne: Sustainable Prototyping of Finite Element Models - Software Release.
\url{https://doi.org/10.5281/zenodo.15789249}
\end{botherref}
\endbibitem

\bibitem[\protect\citeauthoryear{Alk{\"a}mper et~al.}{2016}]{alugrid:16}
\begin{barticle}
\bauthor{\bsnm{Alk{\"a}mper}, \binits{M.}},
\bauthor{\bsnm{Dedner}, \binits{A.}},
\bauthor{\bsnm{Kl{\"o}fkorn}, \binits{R.}},
\bauthor{\bsnm{Nolte}, \binits{M.}}:
\batitle{{The DUNE-ALUGrid Module.}}
\bjtitle{Archive of Numerical Software}
\bvolume{4}(\bissue{1}),
\bfpage{1}--\blpage{28}
(\byear{2016})
\doiurl{10.11588/ans.2016.1.23252}
\end{barticle}
\endbibitem

\bibitem[\protect\citeauthoryear{Le~Bars and
  Worster}{2006}]{lebarsInterfacialConditionsPure2006}
\begin{barticle}
\bauthor{\bsnm{Le~Bars}, \binits{M.}},
\bauthor{\bsnm{Worster}, \binits{M.G.}}:
\batitle{Interfacial conditions between a pure fluid and a porous medium:
  Implications for binary alloy solidification}.
\bjtitle{Journal of Fluid Mechanics}
\bvolume{550}(\bissue{-1}),
\bfpage{149}
(\byear{2006})
\doiurl{10.1017/S0022112005007998}
\end{barticle}
\endbibitem

\bibitem[\protect\citeauthoryear{Swaminathan and
  Voller}{1993}]{swaminathanENTHALPYMETHOD1993}
\begin{barticle}
\bauthor{\bsnm{Swaminathan}, \binits{C.R.}},
\bauthor{\bsnm{Voller}, \binits{V.R.}}:
\batitle{{{On the enthalpy method}}}.
\bjtitle{International Journal of Numerical Methods for Heat \& Fluid Flow}
\bvolume{3}(\bissue{3}),
\bfpage{233}--\blpage{244}
(\byear{1993})
\doiurl{10.1108/eb017528}
\end{barticle}
\endbibitem

\bibitem[\protect\citeauthoryear{Faden
  et~al.}{2019}]{fadenOptimumEnthalpyApproach2019}
\begin{barticle}
\bauthor{\bsnm{Faden}, \binits{M.}},
\bauthor{\bsnm{{K{\"o}nig-Haagen}}, \binits{A.}},
\bauthor{\bsnm{Br{\"u}ggemann}, \binits{D.}}:
\batitle{An {{Optimum Enthalpy Approach}} for {{Melting}} and
  {{Solidification}} with {{Volume Change}}}.
\bjtitle{Energies}
\bvolume{12}(\bissue{5}),
\bfpage{868}
(\byear{2019})
\doiurl{10.3390/en12050868}
\end{barticle}
\endbibitem

\bibitem[\protect\citeauthoryear{Krabbenhoft
  et~al.}{2006}]{krabbenhoftImplicitMixedEnthalpy2006}
\begin{barticle}
\bauthor{\bsnm{Krabbenhoft}, \binits{K.}},
\bauthor{\bsnm{Damkilde}, \binits{L.}},
\bauthor{\bsnm{Nazem}, \binits{M.}}:
\batitle{An implicit mixed enthalpy--temperature method for phase-change
  problems}.
\bjtitle{Heat and Mass Transfer}
\bvolume{43}(\bissue{3}),
\bfpage{233}--\blpage{241}
(\byear{2006})
\doiurl{10.1007/s00231-006-0090-1}
\end{barticle}
\endbibitem

\bibitem[\protect\citeauthoryear{Nedjar}{2002}]{nedjarEnthalpybasedFiniteElement2002}
\begin{barticle}
\bauthor{\bsnm{Nedjar}, \binits{B.}}:
\batitle{An enthalpy-based finite element method for nonlinear heat problems
  involving phase change}.
\bjtitle{Computers \& Structures}
\bvolume{80}(\bissue{1}),
\bfpage{9}--\blpage{21}
(\byear{2002})
\doiurl{10.1016/S0045-7949(01)00165-1}
\end{barticle}
\endbibitem

\bibitem[\protect\citeauthoryear{Terschanski
  et~al.}{2025}]{bryne_simulation_data}
\begin{botherref}
\oauthor{\bsnm{Terschanski}, \binits{B.}},
\oauthor{\bsnm{Klöfkorn}, \binits{R.}},
\oauthor{\bsnm{Dedner}, \binits{A.}},
\oauthor{\bsnm{Kowalski}, \binits{J.}}:
Bryne: Sustainable Prototyping of Finite Element Models - Simulation Data.
\url{https://doi.org/10.5281/zenodo.15850111}
\end{botherref}
\endbibitem

\bibitem[\protect\citeauthoryear{V{\'e}lez
  et~al.}{2015}]{velezTemperaturedependentThermalProperties2015}
\begin{barticle}
\bauthor{\bsnm{V{\'e}lez}, \binits{C.}},
\bauthor{\bsnm{Khayet}, \binits{M.}},
\bauthor{\bsnm{Ortiz De~Z{\'a}rate}, \binits{J.M.}}:
\batitle{Temperature-dependent thermal properties of solid/liquid phase change
  even-numbered n-alkanes: N-{{Hexadecane}}, n-octadecane and n-eicosane}.
\bjtitle{Applied Energy}
\bvolume{143},
\bfpage{383}--\blpage{394}
(\byear{2015})
\doiurl{10.1016/j.apenergy.2015.01.054}
\end{barticle}
\endbibitem

\bibitem[\protect\citeauthoryear{Zhang
  et~al.}{2010}]{zhangOverviewPhaseChange2010}
\begin{barticle}
\bauthor{\bsnm{Zhang}, \binits{P.}},
\bauthor{\bsnm{Ma}, \binits{Z.W.}},
\bauthor{\bsnm{Wang}, \binits{R.Z.}}:
\batitle{An overview of phase change material slurries: {{MPCS}} and {{CHS}}}.
\bjtitle{Renewable and Sustainable Energy Reviews}
\bvolume{14}(\bissue{2}),
\bfpage{598}--\blpage{614}
(\byear{2010})
\doiurl{10.1016/j.rser.2009.08.015}
\end{barticle}
\endbibitem

\bibitem[\protect\citeauthoryear{Chourdakis et~al.}{2022}]{preCICEv2}
\begin{botherref}
\oauthor{\bsnm{Chourdakis}, \binits{G.}},
\oauthor{\bsnm{Davis}, \binits{K.}},
\oauthor{\bsnm{Rodenberg}, \binits{B.}},
\oauthor{\bsnm{Schulte}, \binits{M.}},
\oauthor{\bsnm{Simonis}, \binits{F.}},
\oauthor{\bsnm{Uekermann}, \binits{B.}},
\oauthor{\bsnm{Abrams}, \binits{G.}},
\oauthor{\bsnm{Bungartz}, \binits{H.}},
\oauthor{\bsnm{Cheung~Yau}, \binits{L.}},
\oauthor{\bsnm{Desai}, \binits{I.}},
\oauthor{\bsnm{Eder}, \binits{K.}},
\oauthor{\bsnm{Hertrich}, \binits{R.}},
\oauthor{\bsnm{Lindner}, \binits{F.}},
\oauthor{\bsnm{Rusch}, \binits{A.}},
\oauthor{\bsnm{Sashko}, \binits{D.}},
\oauthor{\bsnm{Schneider}, \binits{D.}},
\oauthor{\bsnm{Totounferoush}, \binits{A.}},
\oauthor{\bsnm{Volland}, \binits{D.}},
\oauthor{\bsnm{Vollmer}, \binits{P.}},
\oauthor{\bsnm{Koseomur}, \binits{O.}}:
{preCICE} v2: A sustainable and user-friendly coupling library [version 2; peer
  review: 2 approved].
Open Research Europe
\textbf{2}(51)
(2022)
\doiurl{10.12688/openreseurope.14445.2}
\end{botherref}
\endbibitem

\bibitem[\protect\citeauthoryear{Rubenstein}{1971}]{stefanProblemRubenstein}
\begin{bbook}
\bauthor{\bsnm{Rubenstein}, \binits{L.I.}}:
\bbtitle{The Stefan Problem / L. I. Rubenstein}.
\bsertitle{Translations of mathematical monographs 27}.
\bpublisher{American Math. Soc.},
\blocation{Providence, RI}
(\byear{1971})
\end{bbook}
\endbibitem

\bibitem[\protect\citeauthoryear{Voller and
  Cross}{1981}]{vollerAccurateSolutionsMoving1981a}
\begin{barticle}
\bauthor{\bsnm{Voller}, \binits{V.}},
\bauthor{\bsnm{Cross}, \binits{M.}}:
\batitle{Accurate solutions of moving boundary problems using the enthalpy
  method}.
\bjtitle{International Journal of Heat and Mass Transfer}
\bvolume{24}(\bissue{3}),
\bfpage{545}--\blpage{556}
(\byear{1981})
\doiurl{10.1016/0017-9310(81)90062-4}
\end{barticle}
\endbibitem

\end{thebibliography}

\end{document}